\documentclass[letterpaper, 10 pt, conference]{ieeeconf}  

\IEEEoverridecommandlockouts                              

\usepackage{amsmath}
\usepackage{amssymb}

\usepackage[pdftex]{graphicx}
\usepackage{subfig}
\graphicspath{{./images/}}
\DeclareGraphicsExtensions{.pdf,.jpeg,.png}

\usepackage{booktabs}
\usepackage{tikz}
\usetikzlibrary{shapes.misc, positioning}

\usepackage{ntheorem}

\newcommand{\COS}[1]{\,\mathrm{c}_{#1}}
\newcommand{\SIN}[1]{\,\mathrm{s}_{#1}}

\title{\LARGE \bf A feedback linearisation algorithm for single-track models\\ with structural stability properties}

\author{Luca Bascetta, Marcello Farina, Alessandro Gabrielli and Matteo Matteucci
	\thanks{L. Bascetta, M. Farina, A. Gabrielli and M. Matteucci are with the Dipartimento di Elettronica, Informazione e Bioingegneria, Politecnico di Milano, Milano, Italy.
		{\tt\small \{luca.bascetta, marcello.farina, alessandro.gabrielli, matteo.matteucci\}@polimi.it}}%
}

\begin{document}
\maketitle
\thispagestyle{empty}
\pagestyle{empty}
	
\begin{abstract}
This paper proposes a feedback linearising law for single-track dynamic models, allowing the design of a trajectory tracking controller exploiting linear control theory. The main characteristics of this algorithm are its simplicity, its independence from any vehicle model parameter, apart from the position of the center of mass, and its robustness. In particular, a numerical bifurcation analysis demonstrates that, for physically meaningful values of the center of mass deviation, the equilibrium is structurally asymptotically stable. Experimental results, concerning the linearising law and its application as inner loop of a trajectory tracking controller, are also presented, confirming the effectiveness of the proposal.
\end{abstract}

\section{Introduction} \label{sec:introduction}
In the last decade, the popularity of research on regulation and trajectory tracking control for mobile robots and, in particular, for autonomous vehicles, has been increasing, and a huge amount of different approaches have been devised. However, the literature still lacks a generic, simple, and flexible control methodology, like the ones available for robotic manipulators since the 80s.
The main reason that hampers the development of such a generic methodology is the complexity introduced by nonholonomic constraints, which make the control problem for nonholonomic mobile robots~\cite{bib:Brockett1983}, for both kinematic and dynamic models, far more complex with respect to controlling omnidirectional mobile robots or manipulators.

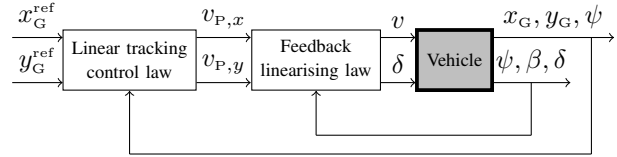
\begin{figure}[htbp]
\tikzstyle{outer_controller} = [draw, rectangle, minimum height=8mm, minimum
width=17mm,align=center]
\tikzstyle{inner_controller} = [draw, rectangle, minimum height=8mm, minimum
width=17mm,align=center]
\tikzstyle{model} = [draw, rectangle, minimum height=8mm, minimum
width=10mm,align=center,line width=0.5mm,fill=lightgray]
\centering
\begin{tikzpicture}
 \node[outer_controller] (a) at (-2.44,0) {\scriptsize Linear tracking\\[-0.1cm] \scriptsize
control law};
 \node[inner_controller] (b) at (0.04,0) {\scriptsize Feedback\\[-0.1cm] \scriptsize
linearising law};
 \node[model] (c) at (1.88,0) {\scriptsize Vehicle};
 \draw[->] (-4.0,0.3) -- (-3.33,0.3) node[above,pos=0.5] {$x_{{\rm\scriptscriptstyle G}}^{{\rm\scriptscriptstyle ref}}$};
 \draw[->] (-4.0,-0.3) -- (-3.33,-0.3) node[above,pos=0.5] {$y_{{\rm\scriptscriptstyle G}}^{{\rm\scriptscriptstyle ref}}$};
 \draw[->] (-1.55,0.3) -- (-0.83,0.3) node[above,pos=0.5] {$v_{{\rm\scriptscriptstyle P},x}$};
 \draw[->] (-1.55,-0.3) -- (-0.83,-0.3) node[above,pos=0.5] {$v_{{\rm\scriptscriptstyle P},y}$};
 \draw[->] (0.9,0.3) -- (1.36,0.3) node[above,pos=0.5] {$v$};
 \draw[->] (0.9,-0.3) -- (1.36,-0.3) node[above,pos=0.5] {$\delta$};
 \draw[->] (2.38,-0.3) -- (3.42,-0.3) node[above,pos=0.5] {$\psi,\beta,\delta$};
 \draw[->] (2.38,0.3) -- (4.0,0.3) node[above,pos=0.5] {$x_{{\rm\scriptscriptstyle G}},y_{{\rm\scriptscriptstyle G}},\psi$};
 \draw[-] (2.9,-0.3) -- (2.9,-1.0);
 \draw[-] (2.9,-1.0) -- (0.04,-1.0);
 \draw[->] (0.04,-1.0) -- (0.04,-0.41);
 \draw[-] (3.7,0.3) -- (3.7,-1.25);
 \draw[-] (3.7,-1.25) -- (-2.44,-1.25);
 \draw[->] (-2.44,-1.25) -- (-2.44,-0.41);
\end{tikzpicture}
\caption{Trajectory tracking control architecture.\label{fig:trajtrack_scheme}}
\end{figure}

Feedback linearisation consists of a nonlinear variable transformation that allows to enforce a linear behaviour to system dynamics. When applicable, it is a promising approach to address the aforementioned problem. It has the advantage of making the design of, e.g., trajectory tracking controllers very flexible, as linear analysis and design tools can be applied without restrictions (Figure~\ref{fig:trajtrack_scheme}). In case of advanced control approaches like Model Predictive Control ({MPC}) and optimal control, the advantage is at least twofold: first, it is easier to design a control algorithm with guaranteed properties, which is more challenging in the nonlinear case (e.g., design of terminal costs and constraints); secondly, linear MPC implementations are less computationally demanding.

The application of feedback linearisation to the field of autonomous (aerial, ground, underwater, or surface) vehicle control has been partly explored and different approaches have been proposed, that can be divided into two main classes. First, linearisation laws tailored to a specific vehicle model exist. For example,~\cite{bib:CaraccioloEtAl1999} considers the dynamic model of a skid-steering vehicle,~\cite{bib:Mistler2001} considers a 4-rotor helicopter,~\cite{bib:OrioloEtAl2002} and~\cite{bib:BardaroEtAl2018} consider a unicycle kinematic model,~\cite{bib:DeLucaEtAl2010} considers the kinematic model of a nonholonomic mobile manipulator,~\cite{bib:Fossen2011} considers the dynamic model of underwater vehicles, and~\cite{bib:Morin2006,bib:Morin2008} present a thorough survey of feedback linearising techniques for unicycle-like and car-like kinematic models of vehicles. However, none of the latter works considers a single-track dynamic model.
Secondly, a generic approach applicable to mechanical systems modelled using the Lagrangian formulation~\cite{bib:CaraccioloEtAl1999}, which includes the single-track dynamic model, exists as well. This approach consists of a partially linearising state feedback, reducing the dynamic model to a purely kinematic one, and on the selection of a particular output that allows to further reduce the resulting (nonlinear) kinematic model to two independent input-output chains of integrators, by way of a fully linearising dynamic feedback.
However, this approach is affected by critical issues that hamper its practical applicability, in particular to single-track models.
First, in order to recast the classical single-track model into the standard form obtained using the Lagrangian formulation, some simplifications have to be enforced, the most common one being a simplification of the tyre model~\cite{bib:CaraccioloEtAl1999,bib:RuccoEtAl2010}, assuming that lateral forces do not depend on vehicle states (i.e., on wheel slip angles). Secondly, the partially linearising state feedback, reducing the dynamic model into a purely kinematic one, is strongly based on model parameters, e.g., the generalised inertia matrix, that are usually uncertain, forcing the introduction of a further robustifying feedback, as it happens for the inverse-dynamics control of manipulators~\cite{bib:SciaviccoSiciliano-2009,bib:BascettaEtAl2010,bib:SpongEtAl-2019}. Thirdly, the partially linearising state feedback, reducing the dynamic model into a purely kinematic one, yields a law whose control variable is the vector of generalised torques that, in the case of a rear-wheel drive single-track model, are the traction torque on the rear wheel and the steering torque. However, in real applications, it is more common to have a cascaded control architecture, where torques are used by the inner velocity loops and the control variables available for the linearising feedback are velocities or positions.

This paper proposes a novel feedback linearisation algorithm for single-track dynamic models that is simple and robust, and does not depend on any model parameter, apart from the position of the center of mass. Assuming the latter is uncertain in a real vehicle, as it can vary statically, due to a different loading condition, or dynamically, due to load transfer phenomena, a numerical analysis has been performed using the MatCont tool~\cite{bib:DhoogeEtAl2008}. This analysis demonstrates that for physically meaningful values of the center of mass deviation, the asymptotic stability of the equilibrium condition is always guaranteed.

The paper is organised as follows. Section~\ref{sec:exp-setup} describes the experimental setup later used to test the proposed control law. Section~\ref{sec:model+FB} introduces the single-track dynamic model and the feedback linearising law, including the numerical stability analysis. Section~\ref{sec:exp-results} reports the results of the experiments conducted on the feedback linearising law only, and on a trajectory tracking controller whose inner loop is constituted by the proposed linearising law. Finally, conclusions are drawn in Section~\ref{sec:conclusion}.

\section{The experimental setup} \label{sec:exp-setup}
\begin{figure}[htbp]
	\begin{center}
		\includegraphics[width=0.7\columnwidth]{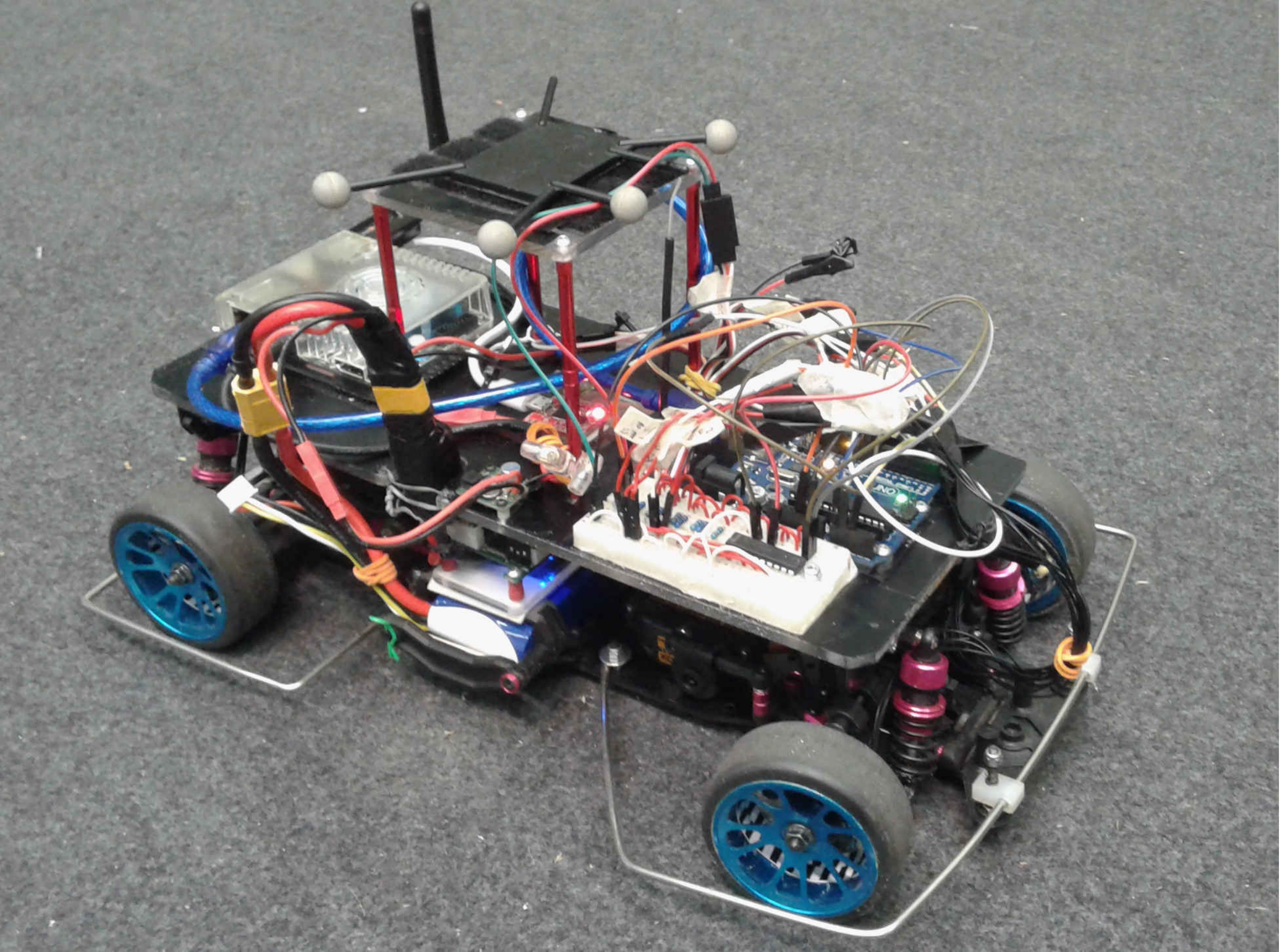}
		\caption{The experimental platform.\label{fig:exp_platform}}
	\end{center}
\end{figure}

For numerical analysis and experimental validation, a 1:10 scale car-like vehicle (Figure~\ref{fig:exp_platform}), inspired by the ones used by {ETH} and Georgia Institute of Technology researchers~\cite{bib:Jelavic2017,bib:Williams2016,bib:Liniger2015}, has been considered. The platform is a rear-wheel drive car, actuated by a brushless motor and equipped with four independent suspensions and an electric steering servo. The car can autonomously drive thanks to the installation of:
\begin{itemize}
\item a computation unit Odroid {XU4}, that runs a ROS architecture, including the controller that is implemented in {C++} as a {ROS} node, having a rate of $100\,\mathrm{Hz}$;
\item an Arduino {UNO} board, that provides a bidirectional communication with actuators.
\end{itemize}
The car is equipped with an {IMU}, providing linear acceleration, angular velocity, and attitude measurements, and with a marker that allows to track vehicle position and orientation at a frequency of $100\,\mathrm{Hz}$ using a 12-camera OptiTrack motion tracking system. These data are used to compute vehicle position, yaw and sideslip angles.
Finally, a four square meter carpet is used to simulate the road surface, ensuring a constant and uniform tyre-ground interaction.\\
Using a car model instead of a real vehicle allows to execute aggressive manoeuvres without running the risk of damaging the car or the environment, or of harming people; as otherwise it could be done only in simulation. The adoption of a scaled vehicle model is also fostered by the existence of a dynamic similitude, expressed by the Buckingham-Pi theorem~\cite{bib:Brennan2001,bib:Hoblet2003}. According to it, the solutions to the nonlinear differential equations modelling a real vehicle are proved to be identical, after accounting for the dimensional scaling of each parameter in the equations, to the solutions to the differential equations describing the scaled model.
%
\section{Single-track model and feedback linearisation} \label{sec:model+FB}
\begin{figure}[htbp]
	\begin{center}
      \includegraphics[width=0.8\columnwidth]{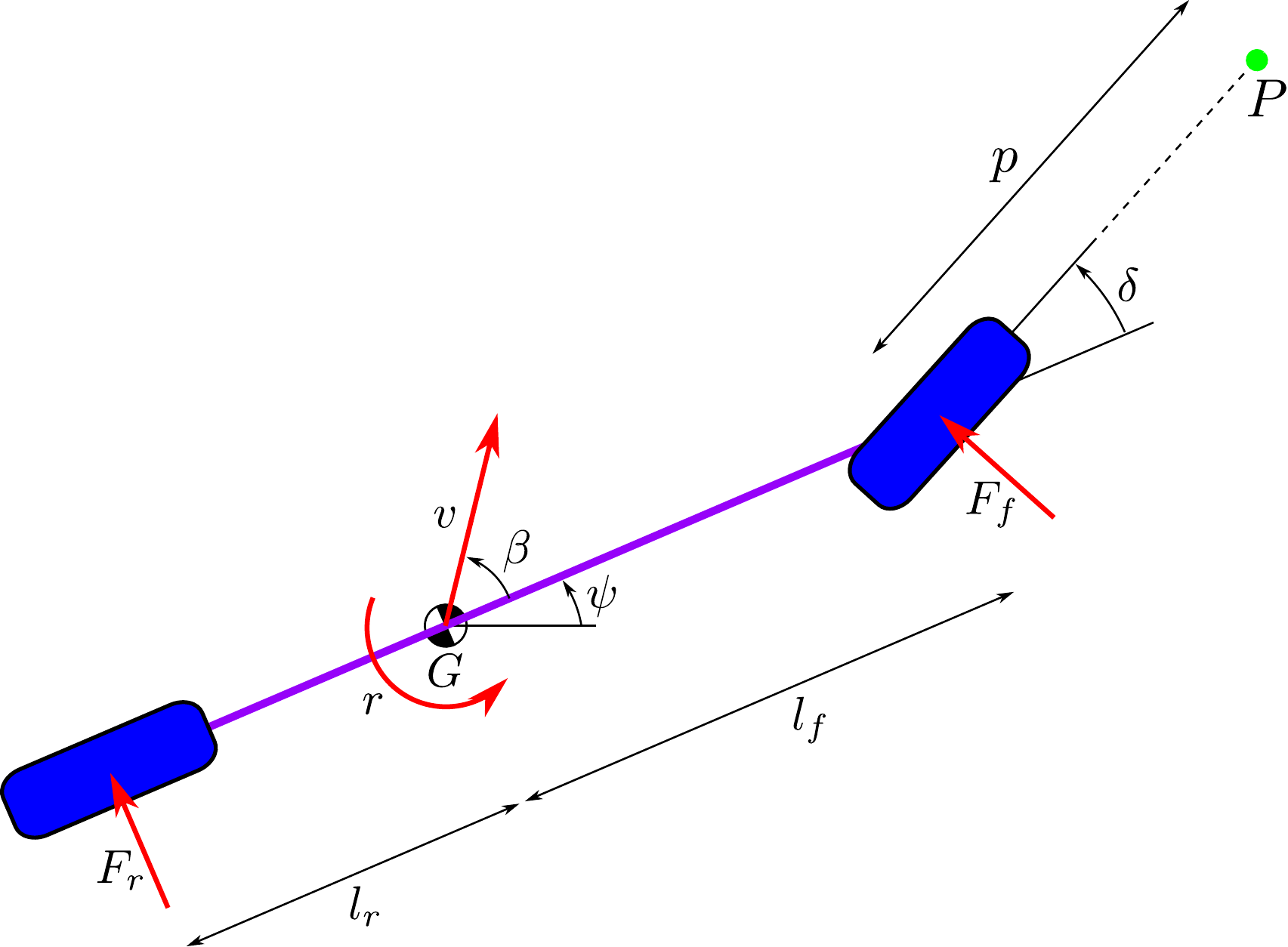}
      \caption{Single-track vehicle model (figure shows the quantities used to derive the motion model).\label{fig:single-track}}
    \end{center}
\end{figure}

As it is common in the control and robotics literature and supported by experimental evidence~\cite{bib:Kong2015}, the longitudinal and lateral dynamics of a vehicle, that are considered as relevant for the design of a trajectory tracking controller, are represented using the single-track approximation (Figure~\ref{fig:single-track}), based on which the motion equations are developed lumping together the wheels on the same front or rear axle at its centreline, and assuming the vehicle mass is concentrated in the center of mass $G$. This approximation is based on the following standard assumptions:
\begin{itemize}
\item ground slope, longitudinal load transfer, and pitching and rolling motions are neglected;
\item a linear tyre model is considered, i.e.,
\begin{displaymath}
F_f=C_f\alpha_f\qquad F_r=C_r\alpha_r
\end{displaymath}
where $F_f$, $F_r$ are front and rear lateral forces, $C_f$, $C_r$ are front and rear cornering stiffness, and $\alpha_f$, $\alpha_r$, front and rear slip angles;
\item sideslip and steering angles are small enough to introduce the approximations\footnote{Here and in the following of the paper, $\SIN{x}$ and $\COS{x}$ are used in place of $\sin(x)$ and $\cos(x)$, respectively.} $\SIN{x}\approx x$ and $\COS{x}\approx 1$;
\item vehicle is rear-wheel drive and braking forces are neglected;
\item vehicle velocity is slowly varying, i.e., the Newton equation related to the longitudinal motion is considered at steady state.
\end{itemize}
\begin{table}
	\centering
	\caption{Experimental Platform Data}
	\label{tab:driftingcar_data}
	\begin{tabular}{lll}
		\toprule
		Mass & $\phantom{15}1.9$ & $\mathrm{Kg}$ \\
		Yaw moment of inertia & $\phantom{15}0.0251$ & $\mathrm{Kg}\,\mathrm{m}^2$ \\
		Distance of center of mass from front axle & $\phantom{15}0.1368$ & $\mathrm{m}$ \\
		Distance of center of mass from rear axle & $\phantom{15}0.1232$ & $\mathrm{m}$\\
		Tyre-ground friction coefficient & $\phantom{15}0.25$ & \\
		Front wheels cornering stiffness & $\phantom{1}58.085$ & $\mathrm{N}\,\mathrm{rad}^{-1}$ \\
		Rear wheels cornering stiffness & $130.805$ & $\mathrm{N}\,\mathrm{rad}^{-1}$\\
		\bottomrule
	\end{tabular}
\end{table}
The single-track motion model is
\begin{equation} \label{eq:singletrack_model}
\begin{aligned}
\dot{\psi}&=r\\
\dot{r}&=\frac{C_r l_r-C_f l_f}{I_z}\beta-\frac{C_f l_f^2+C_r l_r^2}{I_z v}r+\frac{C_f l_f}{I_z}\delta\\
\dot{\beta}&=-\frac{C_f+C_r}{mv}\beta+\left(\frac{C_r l_r-C_f l_f}{m v^2}-1\right)r+\frac{C_f}{mv}\delta
\end{aligned}
\end{equation}
where
\begin{itemize}
\item $\psi$ and $r$ are yaw angle and yaw rate, respectively;
\item $\beta$ is the sideslip angle;
\item $v$ is the vehicle velocity;
\item $\delta$ is the steering angle;
\item $l_f$ and $l_r$ are the distances of the vehicle center of mass $G$ from the front and rear axles, respectively;
\item $I_z$ is the yaw inertia referred to the center of mass;
\item $m$ is the mass.
\end{itemize}
The time evolution of the position $x_{G}$, $y_{G}$ of the center of mass $G$ with respect to an inertial reference frame is given by
\begin{equation} \label{eq:kinematic_model}
\dot{x}_{G}=v\COS{\psi+\beta}\qquad
\dot{y}_{G}=v\SIN{\psi+\beta}
\end{equation}
Note that the state variables of model~\eqref{eq:singletrack_model}-\eqref{eq:kinematic_model} are $\psi$, $r$, $\beta$, $x_G$, and $y_ G$, while $v$ and $\delta$ are its inputs.

\begin{figure}[htbp]
	\begin{center}
		\includegraphics[width=0.56\columnwidth]{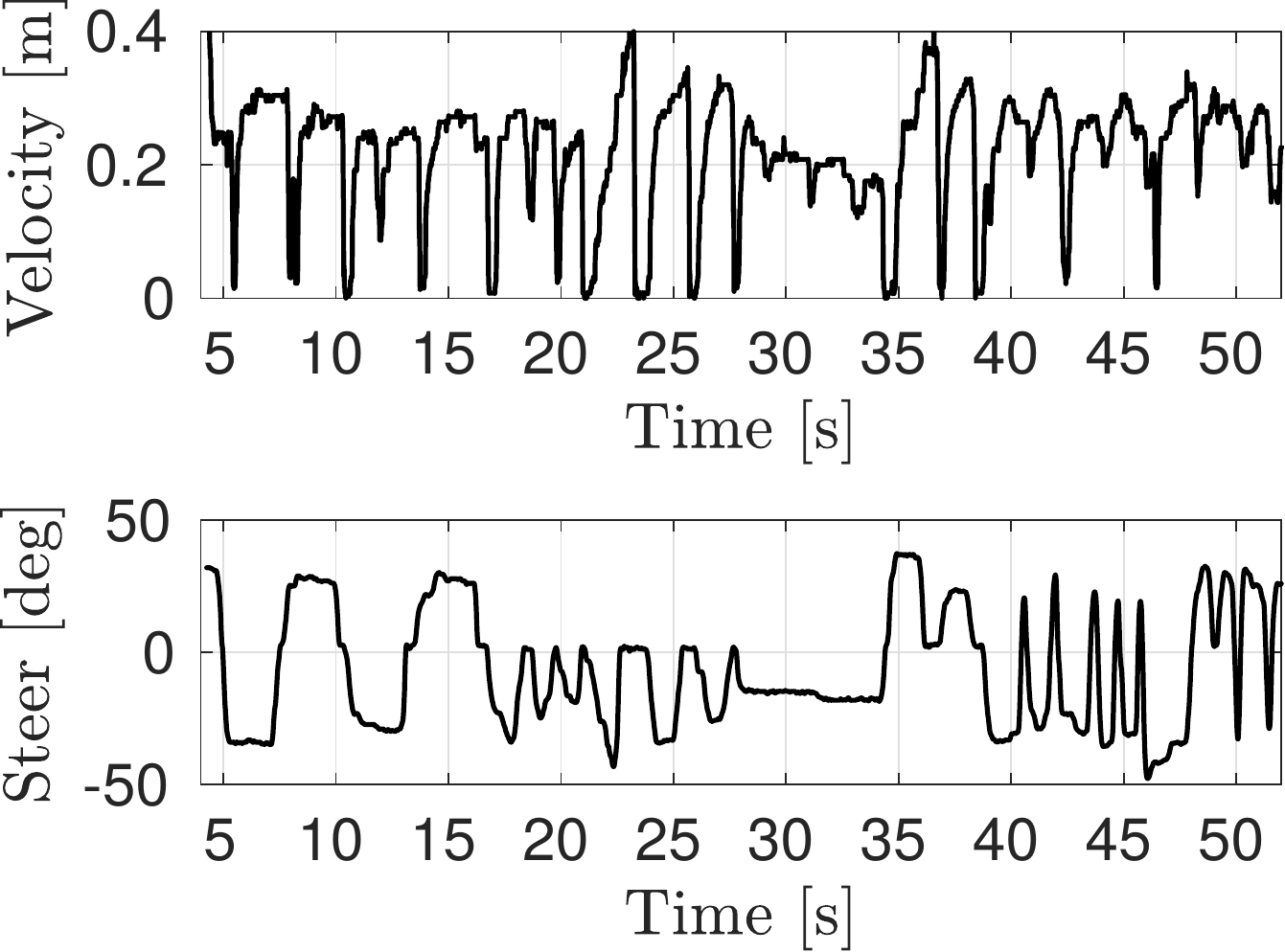}\hfill
		\includegraphics[width=0.41\columnwidth]{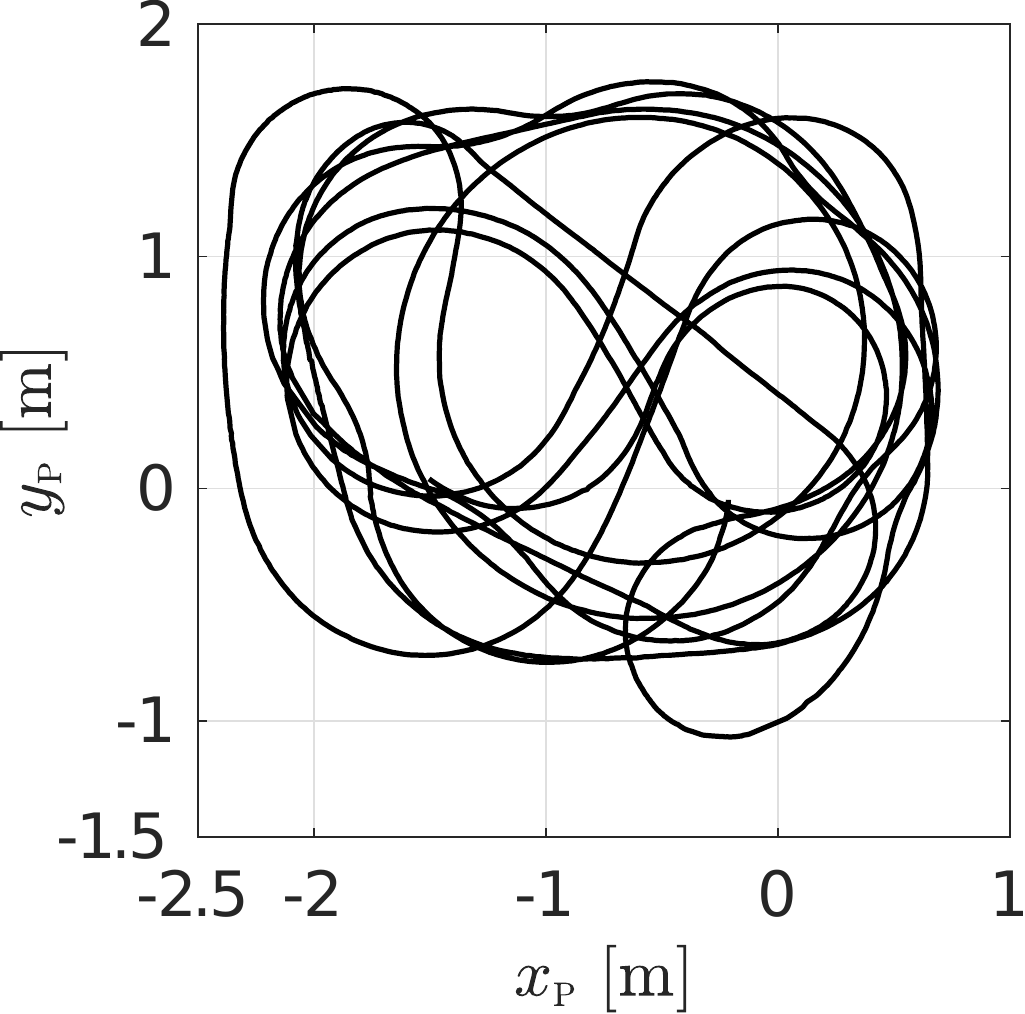}
		\caption{Part of the inputs and vehicle trajectory used for single-track model validation.\label{fig:exp-ident_inputs}}
	\end{center}
\end{figure}

\begin{figure}[htbp]
	\begin{center}
		\includegraphics[width=0.49\columnwidth]{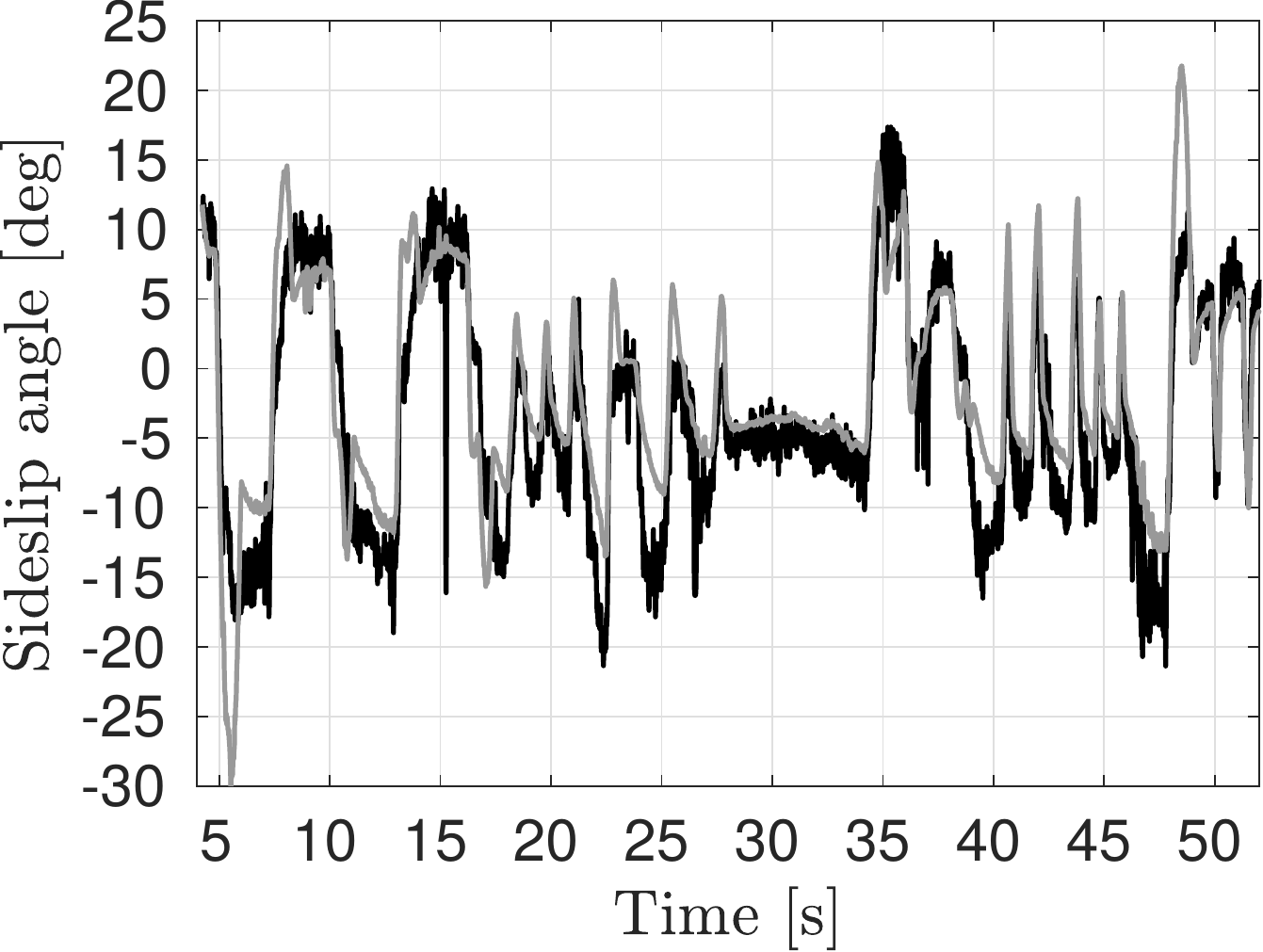}\hfill
		\includegraphics[width=0.452\columnwidth]{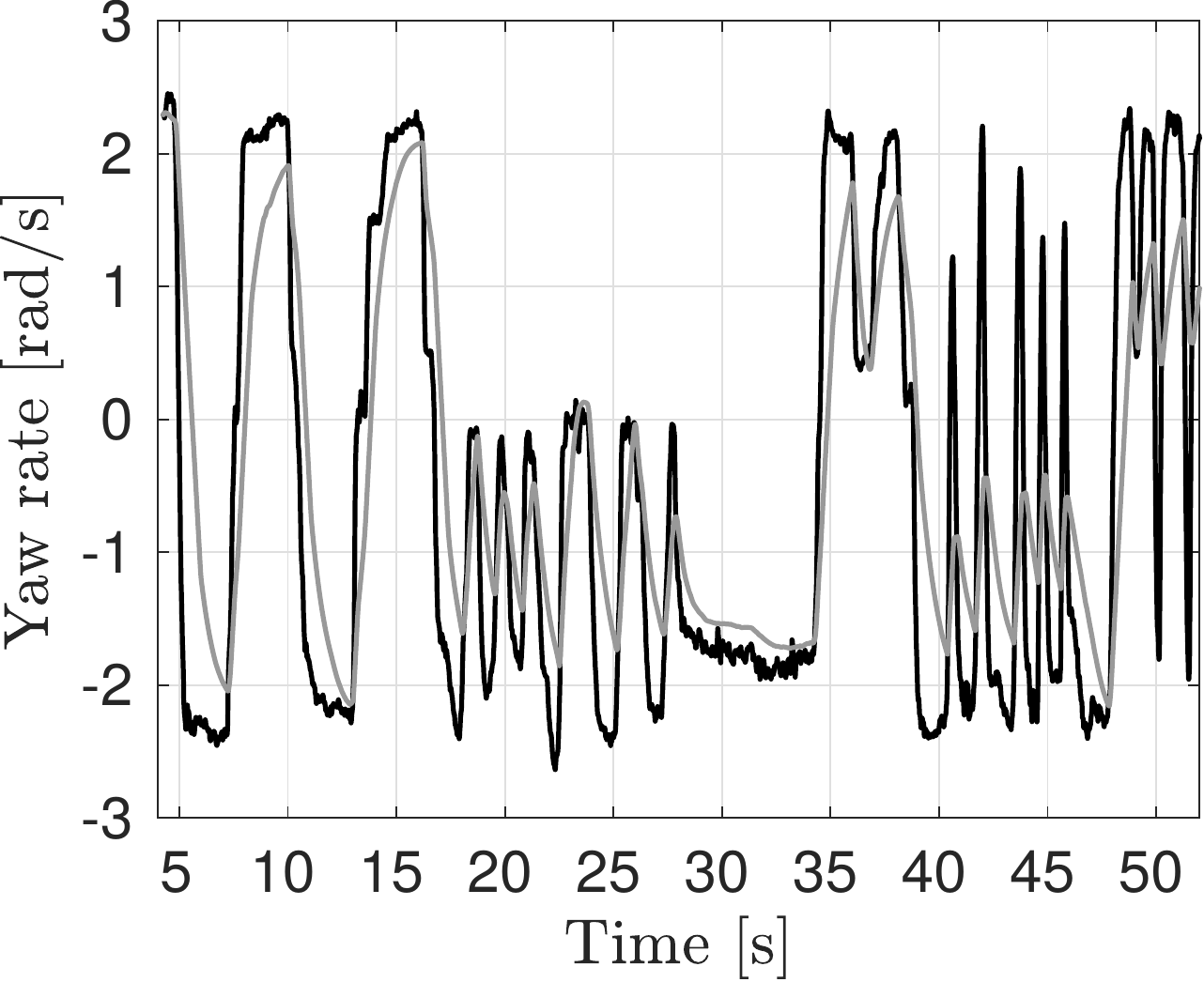}\\
		\includegraphics[width=0.49\columnwidth]{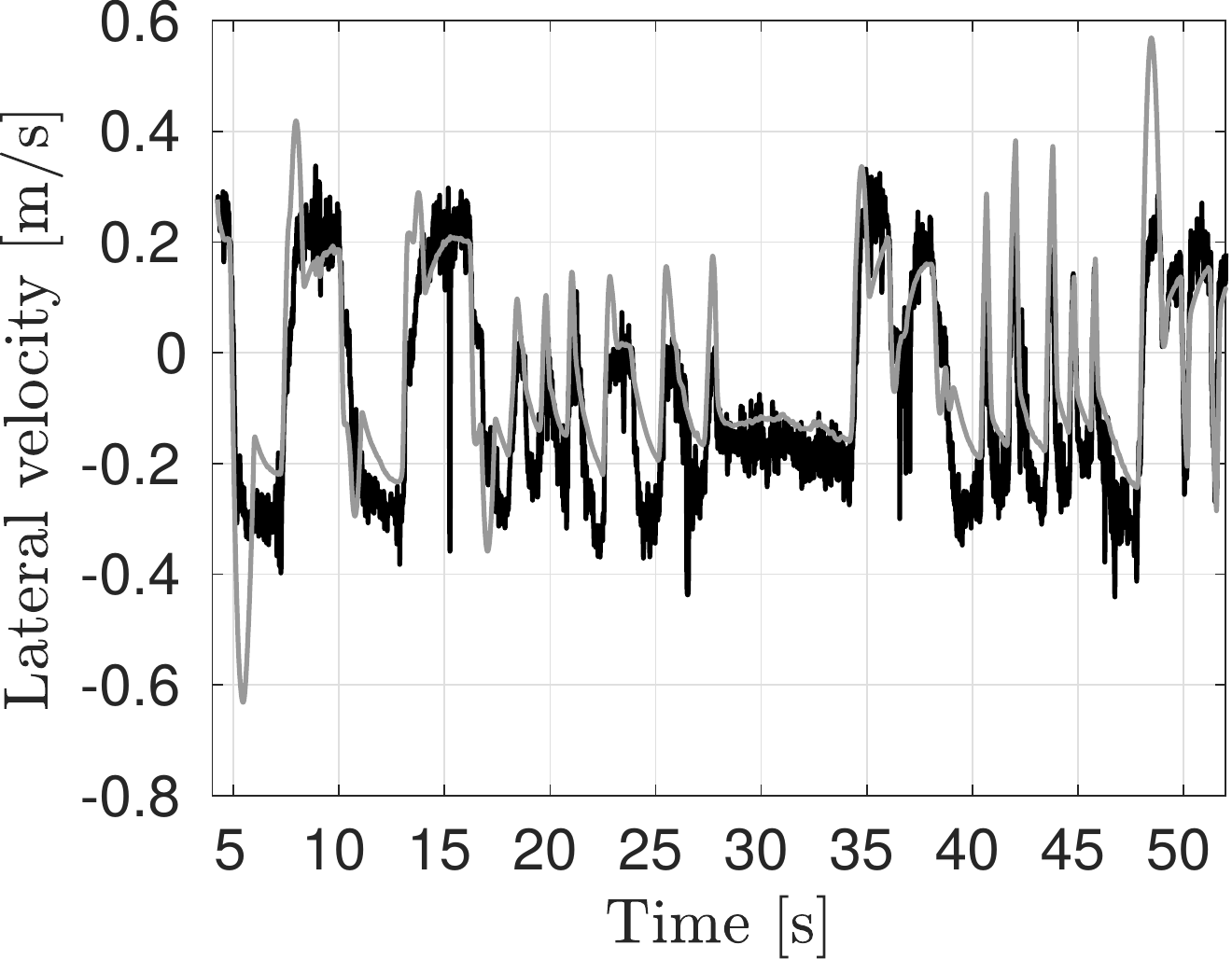}\hfill
		\includegraphics[width=0.475\columnwidth]{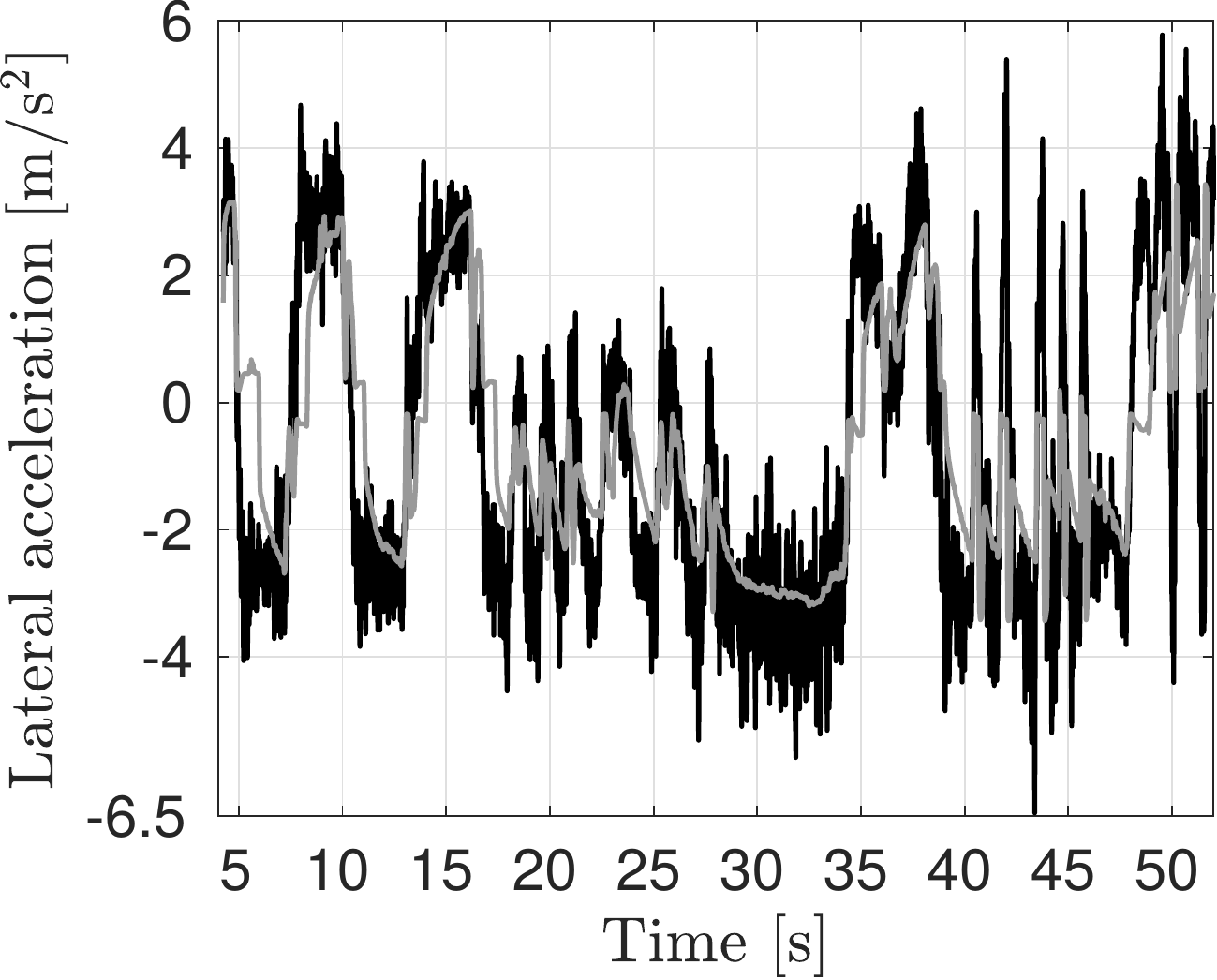}
		\caption{Comparison between experimental data (black line) and single-track simulation (gray line).\label{fig:exp-ident_states}}
	\end{center}
\end{figure}

Concerning the experimental platform, vehicle mass $m$ and center of mass position, i.e., $l_f$ and $l_r$, have been measured with a weight balance, while the other parameters have been identified using the normalised mean prediction error (NMPE) as in~\cite{bib:Fagiano2018}. The main vehicle parameters, resulting from the identification procedure, are reported in Table~\ref{tab:driftingcar_data}.\\
Regarding the validity of the used model, Figures~\ref{fig:exp-ident_inputs} and~\ref{fig:exp-ident_states} report a comparison between experimental data, taken from a validation dataset, and simulated ones. The low value of the NMPE achieved in the identification procedure ($0.58$), and the corresponding good accordance between experimental and simulation data, demonstrate that the single-track model is able to correctly reproduce the relevant vehicle dynamics.

\subsection{Feedback linearising control law}\label{sub:sec:feedback_linearization}
The proposed feedback linearisation procedure is derived with the aim of enforcing the dynamics of a point $\rm{P}$ at distance $p$ from the front wheel axle, in the steering direction\footnote{The position of point $P$ can be arbitrarily selected, excluding the case $p=0$. Intuitively, the closer point $P$ is to the front wheel contact point, the more accurately it describes the motion of the vehicle.} (see again Figure~\ref{fig:single-track}). Its coordinates are ${x}_{{\rm\scriptscriptstyle P}}$ and ${y}_{{\rm\scriptscriptstyle P}}$, where
\begin{equation}  \label{eq:pointP}
\begin{aligned}
{x}_{{\rm\scriptscriptstyle P}}&={x}_{\rm\scriptscriptstyle G}+l_f\COS{\psi}+p\COS{\psi+\delta}\\
{y}_{{\rm\scriptscriptstyle P}}&={y}_{\rm\scriptscriptstyle G}+l_f\SIN{\psi}+p\SIN{\psi+\delta}
\end{aligned}
\end{equation}
Therefore
\begin{equation} \label{eq:pointP_dot}
\begin{aligned}
\dot{x}_{{\rm\scriptscriptstyle P}}&=v\COS{\psi+\beta}-l_f r\SIN{\psi}-p(r+u_{\delta})\SIN{\psi+\delta}\\
\dot{y}_{{\rm\scriptscriptstyle P}}&=
v\SIN{\psi+\beta}+l_f r\COS{\psi}+p(r+u_{\delta})\COS{\psi+\delta}
\end{aligned}
\end{equation}
where $u_{\delta}$ has been used as input to the single-track model in place of $\delta$, without loss of generality, introducing the following additional system equation
\begin{equation} \label{eq:deltadot}
\dot{\delta}=u_{\delta}
\end{equation}
Suitable reference velocities for point $\rm{P}$, i.e., $v_{{\rm\scriptscriptstyle P},x}$ and $v_{{\rm\scriptscriptstyle P},y}$, can be imposed by setting
\begin{equation} \label{eq:double_integratorP}
\dot{x}_{{\rm\scriptscriptstyle P}}=v_{{\rm\scriptscriptstyle P},x}\qquad
\dot{y}_{{\rm\scriptscriptstyle P}}=v_{{\rm\scriptscriptstyle P},y}
\end{equation}
and this can be achieved selecting at each time instant
\begin{equation} \label{eq:feedback_lin}
\begin{aligned}
v&=\frac{v_{{\rm\scriptscriptstyle P},x}\COS{\psi+\delta}+v_{{\rm\scriptscriptstyle P},y}\SIN{\psi+\delta}
-rl_f\SIN{\delta}}{\COS{\beta-\delta}}\\
u_{\delta}&=\frac{v_{{\rm\scriptscriptstyle P},y}\COS{\psi+\beta}-v_{{\rm\scriptscriptstyle P},x}\SIN{\psi+\beta}
-rl_f\COS{\beta}}{p\COS{\beta-\delta}}-r
\end{aligned}
\end{equation}
Note that, in~\eqref{eq:feedback_lin}, a singularity exists in $\beta-\delta=\pi/2+k\pi$, where $k$ is any integer number. This singularity, however, does not limit the practical applicability of the linearising law. In fact, even in the case of a drifting manoeuvre, characterised by sideslip angles up to $\pm 40\,\textrm{deg}$, the relation $\beta-\delta=\pi/2+k\pi$ is never satisfied, at least for values of the steering angle in the typical range of a standard vehicle, i.e., up to $\pm 50\,\textrm{deg}$.

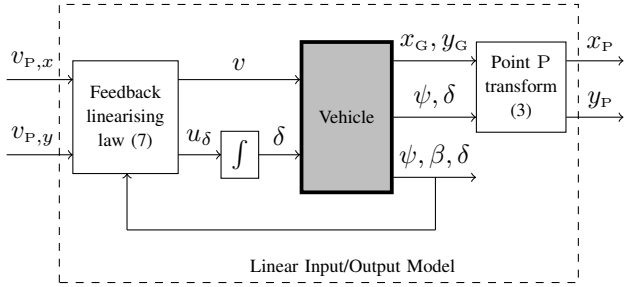
\begin{figure}[htbp]
\tikzstyle{controller} = [draw, rectangle, minimum height=15mm, minimum
width=14mm,align=center]
\tikzstyle{model} = [draw, rectangle, minimum height=20mm, minimum
width=12mm,align=center,line width=0.5mm,fill=lightgray]
\tikzstyle{integrator} = [draw, rectangle, minimum height=5.25mm, minimum
width=5mm,align=center]
\tikzstyle{Ptransform} = [draw, rectangle, minimum height=12mm, minimum
width=8mm,align=center]
\centering
\begin{tikzpicture}
 \node[controller] (a) at (-0.52,0) {\scriptsize Feedback\\[-0.1cm] \scriptsize
linearising\\[-0.1cm] \scriptsize law~\eqref{eq:feedback_lin}};
 \node[integrator] (c) at (1.0,-0.5) {$\int$};
 \node[model] (b) at (2.42,0) {\scriptsize Vehicle};
 \node[Ptransform] (d) at (4.74,0.4) {\scriptsize Point $\rm{P}$\\[-0.1cm] \scriptsize transform\\[-0.1cm] \scriptsize \eqref{eq:pointP}};
 \draw[->] (-2.1,0.5) -- (-1.22,0.5) node[above,pos=0.4] {$v_{{\rm\scriptscriptstyle P},x}$};
 \draw[->] (-2.1,-0.5) -- (-1.22,-0.5) node[above,pos=0.4] {$v_{{\rm\scriptscriptstyle P},y}$};
 \draw[->] (0.18,0.5) -- (1.8,0.5) node[above,pos=0.5] {$v$};
 \draw[->] (0.18,-0.5) -- (0.75,-0.5) node[above,pos=0.5] {$u_{\delta}$};
 \draw[->] (1.25,-0.5) -- (1.8,-0.5) node[above,pos=0.5] {$\delta$};
 \draw[->] (3.04,-0.8) -- (4.14,-0.8) node[above,pos=0.5] {$\psi,\beta,\delta$};
 \draw[->] (3.04,0.75) -- (4.14,0.75) node[above,pos=0.5] {$x_{{\rm\scriptscriptstyle G}},y_{{\rm\scriptscriptstyle G}}$};
 \draw[->] (3.04,0) -- (4.14,0) node[above,pos=0.5] {$\psi,\delta$};
 \draw[->] (5.34,0) -- (6.1,0) node[above,pos=0.6] {$y_{{\rm\scriptscriptstyle P}}$};
 \draw[->] (5.34,0.75) -- (6.1,0.75) node[above,pos=0.6] {$x_{{\rm\scriptscriptstyle P}}$};
 \draw[-] (3.6,-0.8) -- (3.6,-1.5);
 \draw[-] (3.6,-1.5) -- (-0.5,-1.5);
 \draw[->] (-0.5,-1.5) -- (-0.5,-0.75);
 \draw [dashed](-1.4,-2.2) rectangle (5.5,1.5);
 \node at (2.5,-2) {\scriptsize Linear Input/Output Model};
\end{tikzpicture}
\caption{Feedback linearised system.\label{fig:FB_lin_scheme}}
\end{figure}

The comprehensive dynamical system, obtained through the feedback linearisation approach, is sketched in the block diagram of Figure~\ref{fig:FB_lin_scheme}, where the gray box represents the vehicle while the other boxes are the components of the feedback linearising controller.
\subsection{Feedback linearised system} \label{subsec:feedback_linearised system}
Though feedback linearisation is a powerful tool, its practical applicability in a realistic scenario, where a perfect knowledge of the plant parameters can never be assumed, can be hampered by severe robustness issues that, in the worst case, can lead to instability.
An analysis of the feedback linearised system (Figure~\ref{fig:FB_lin_scheme}), in presence of parameter uncertainty, is thus of utmost importance to guarantee the practical applicability of~\eqref{eq:feedback_lin}.
\begin{figure}[htbp]
	\begin{center}
      \includegraphics[width=0.6\columnwidth]{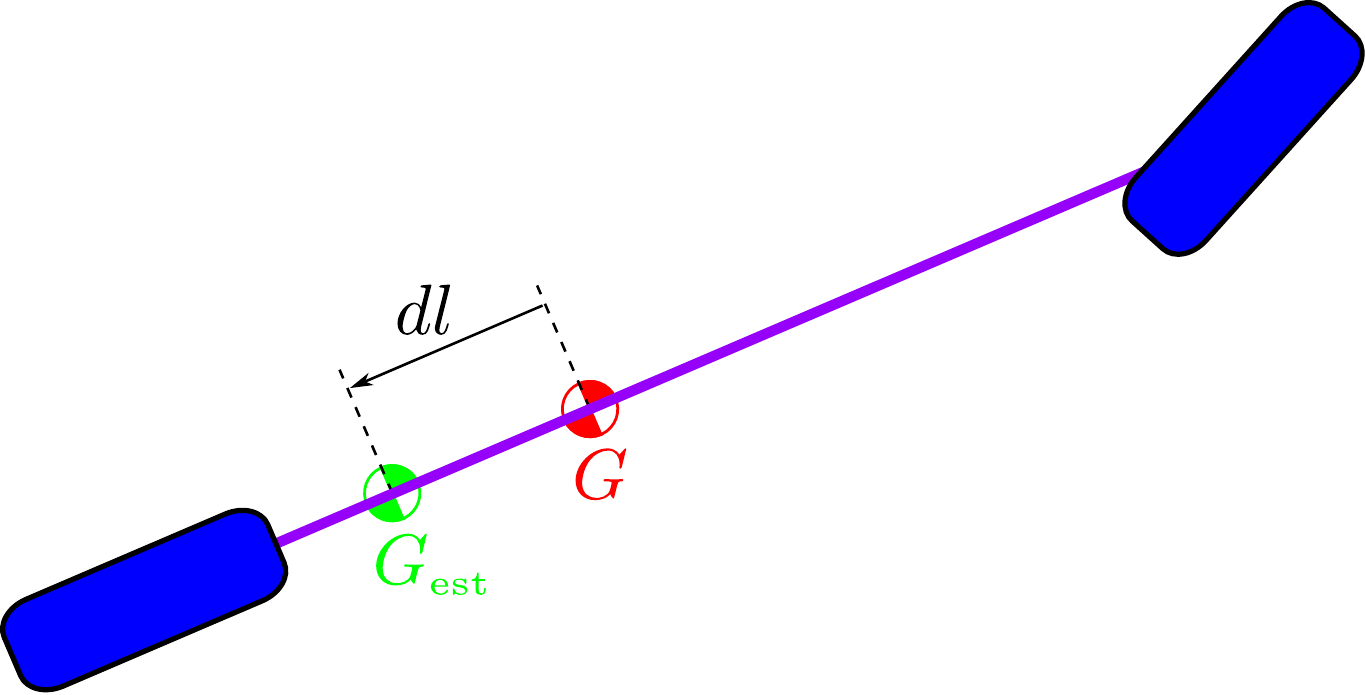}
      \caption{The uncertainty upon the center of mass position.\label{fig:singletrack_uncertainty}}
    \end{center}
\end{figure}

The main advantage of the feedback linearising law \eqref{eq:feedback_lin} lies in its simplicity, as it does not depend on any relevant data but the center of mass position. It is however worthwhile to analyse the structural robustness of the feedback linearised system with respect to possible uncertainties on the position of point G, i.e., to parameter $l_f$.
To this aim, linearising law~\eqref{eq:feedback_lin} is here rewritten replacing $l_f$ with its estimated value $l_f^{\rm\scriptscriptstyle est}$, as follows
\begin{equation} \label{eq:feedback_lin_uncert}
\begin{aligned}
v&=\frac{v_{{\rm\scriptscriptstyle P},x}\COS{\psi+\delta}+v_{{\rm\scriptscriptstyle P},y}\SIN{\psi+\delta}
-rl_f^{\rm\scriptscriptstyle est}\SIN{\delta}}{\COS{\beta-\delta}}\\
u_{\delta}&=\frac{v_{{\rm\scriptscriptstyle P},y}\COS{\psi+\beta}-v_{{\rm\scriptscriptstyle P},x}\SIN{\psi+\beta}
-rl_f^{\rm\scriptscriptstyle est}\COS{\beta}}{p\COS{\beta-\delta}}-r
\end{aligned}
\end{equation}
and the center of mass deviation $dl=l_f^{\rm\scriptscriptstyle est}-l_f$ is introduced as a measure of uncertainty.
To this regard, Figure~\ref{fig:singletrack_uncertainty} shows the possible displacement between the estimated position of the center of mass $G_{\rm\scriptscriptstyle est}$ and the real one $G$. In particular, note that $dl$ is positive if $G_{\rm\scriptscriptstyle est}$ is behind $G$, while it is negative if $G_{\rm\scriptscriptstyle est}$ is ahead of $G$. The range in which $dl$ takes physically meaningful values is, indeed, $[-l_f,l_r]$.\\
Replacing~\eqref{eq:feedback_lin_uncert} in~\eqref{eq:pointP_dot} yields
\begin{equation} \label{eq:P2dot_real}
\dot{x}_{{\rm\scriptscriptstyle P}}=v_{{\rm\scriptscriptstyle P},x}+dl\SIN{\psi}r\qquad
\dot{y}_{{\rm\scriptscriptstyle P}}=v_{{\rm\scriptscriptstyle P},y}-dl\COS{\psi}r
\end{equation}
\subsection{Structural stability analysis}
\label{sec:structural_stab}
\begin{figure}[htbp]
      \centering
      \includegraphics[width=0.8\columnwidth]{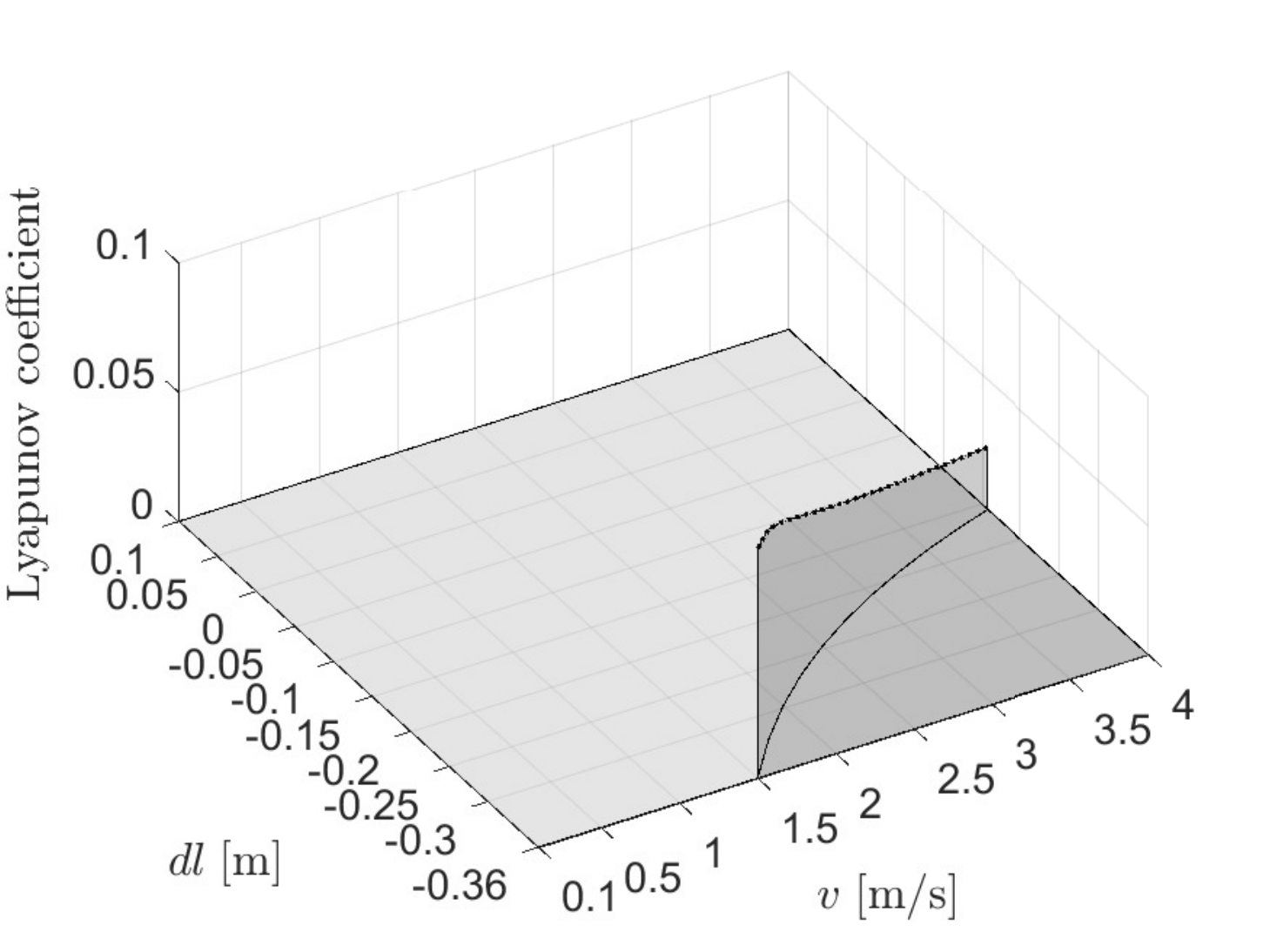}
      \caption{Stability properties of the selected equilibrium point as a function of $\bar{v}$ and $dl$. Light gray region: asymptotic stability; dark gray region: instability; black solid line: Hopf bifurcation points; dotted black line: value of the first Lyapunov coefficient for all bifurcation points.}\label{fig:bifurcaz}
\end{figure}

\begin{figure}[htbp]
	\centering
	\includegraphics[width=0.8\columnwidth]{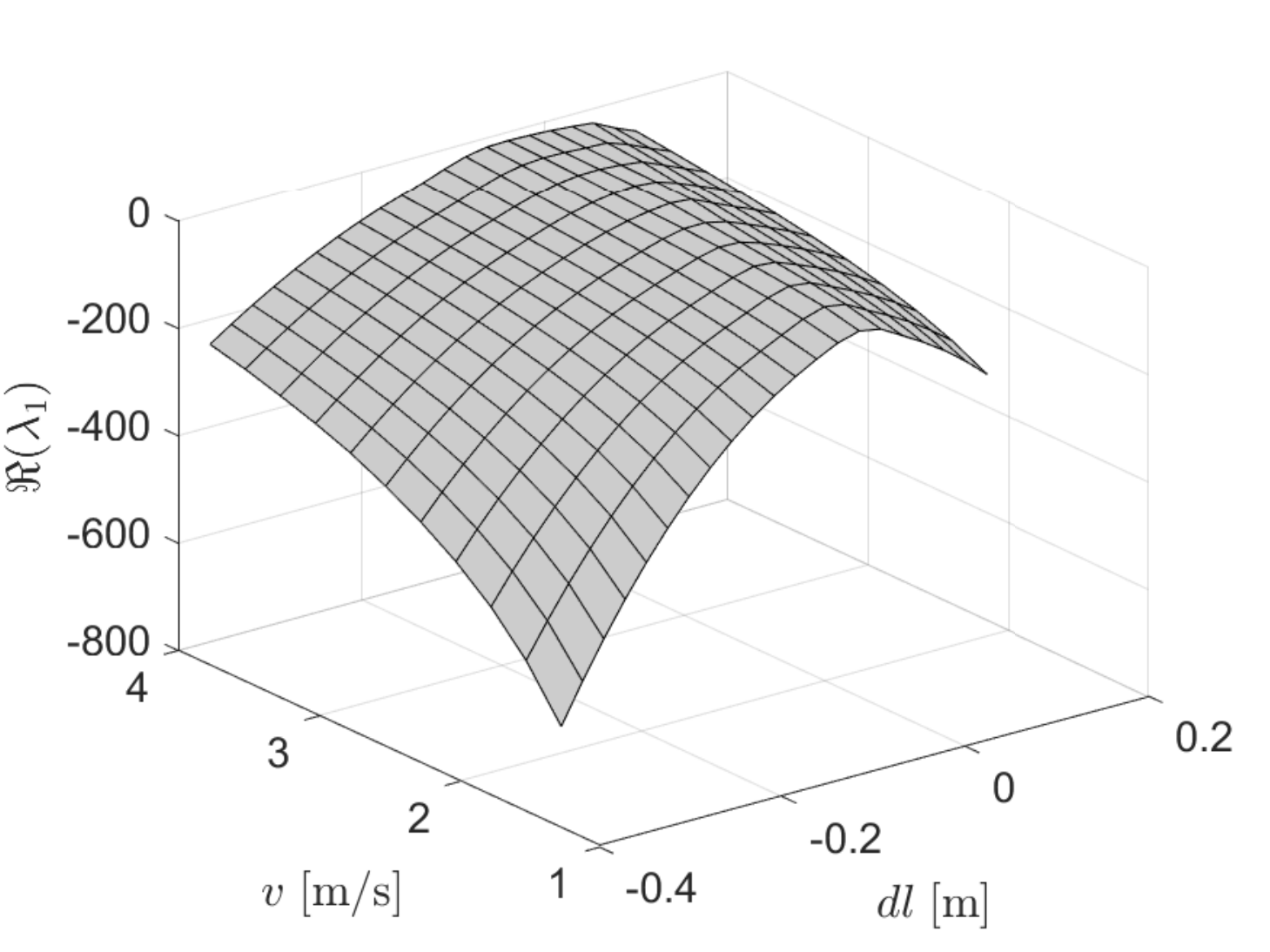}
	\caption{Real part of eigenvalue $\lambda_1$ of the dynamics of $\xi$, linearised around the defined equilibrium value. Light gray is used to define the region in which $\mathcal{R}(\lambda_1)<0$.}\label{fig:l1}
\end{figure}

\begin{figure}[htbp]
	\centering
	\includegraphics[width=0.8\columnwidth]{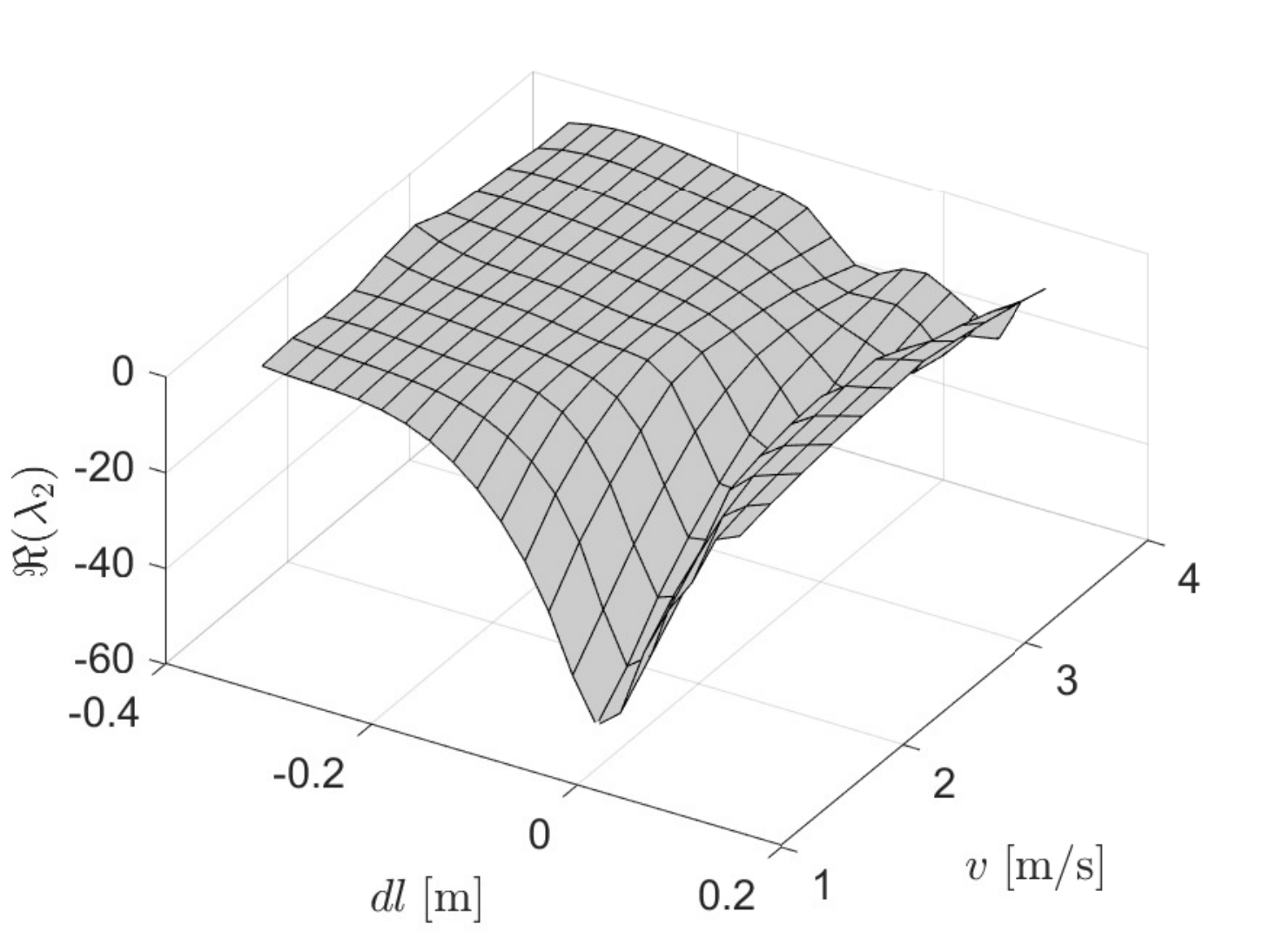}
	\caption{Real part of eigenvalue $\lambda_2$ of the dynamics of $\xi$, linearised around the defined equilibrium value. Light gray is used to define the region in which $\mathcal{R}(\lambda_2)<0$.}\label{fig:l2}
\end{figure}

\begin{figure}[htbp]
	\centering
	\includegraphics[width=0.8\columnwidth]{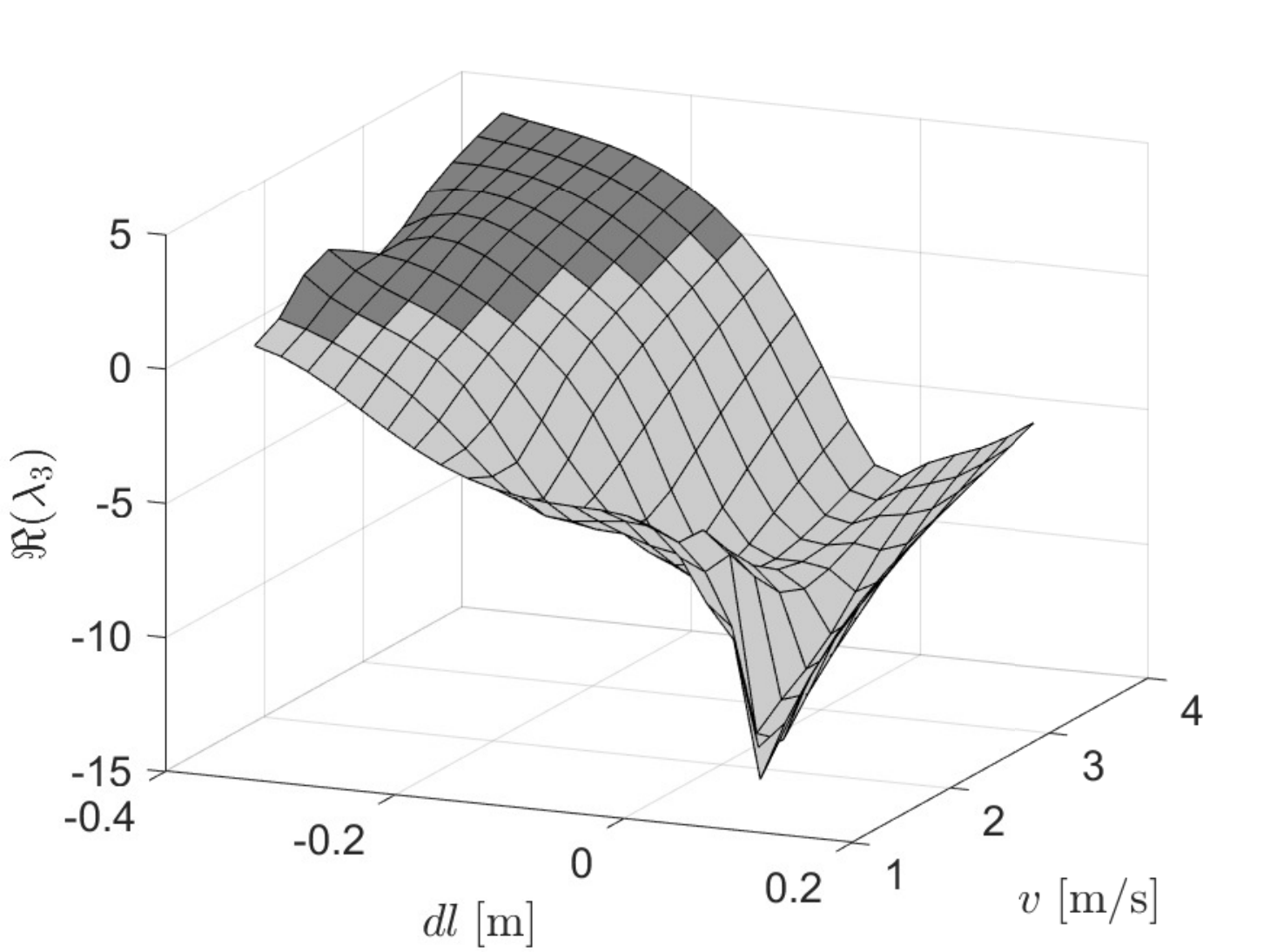}
	\caption{Real part of eigenvalue $\lambda_3$ of the dynamics of $\xi$, linearised around the defined equilibrium value. Light gray is used to define the region in which $\mathcal{R}(\lambda_3)<0$, while dark gray is used to define the region where $\mathcal{R}(\lambda_3)\geq 0$.}\label{fig:l3}
\end{figure}

\begin{figure}[htbp]
	\centering
	\includegraphics[width=0.8\columnwidth]{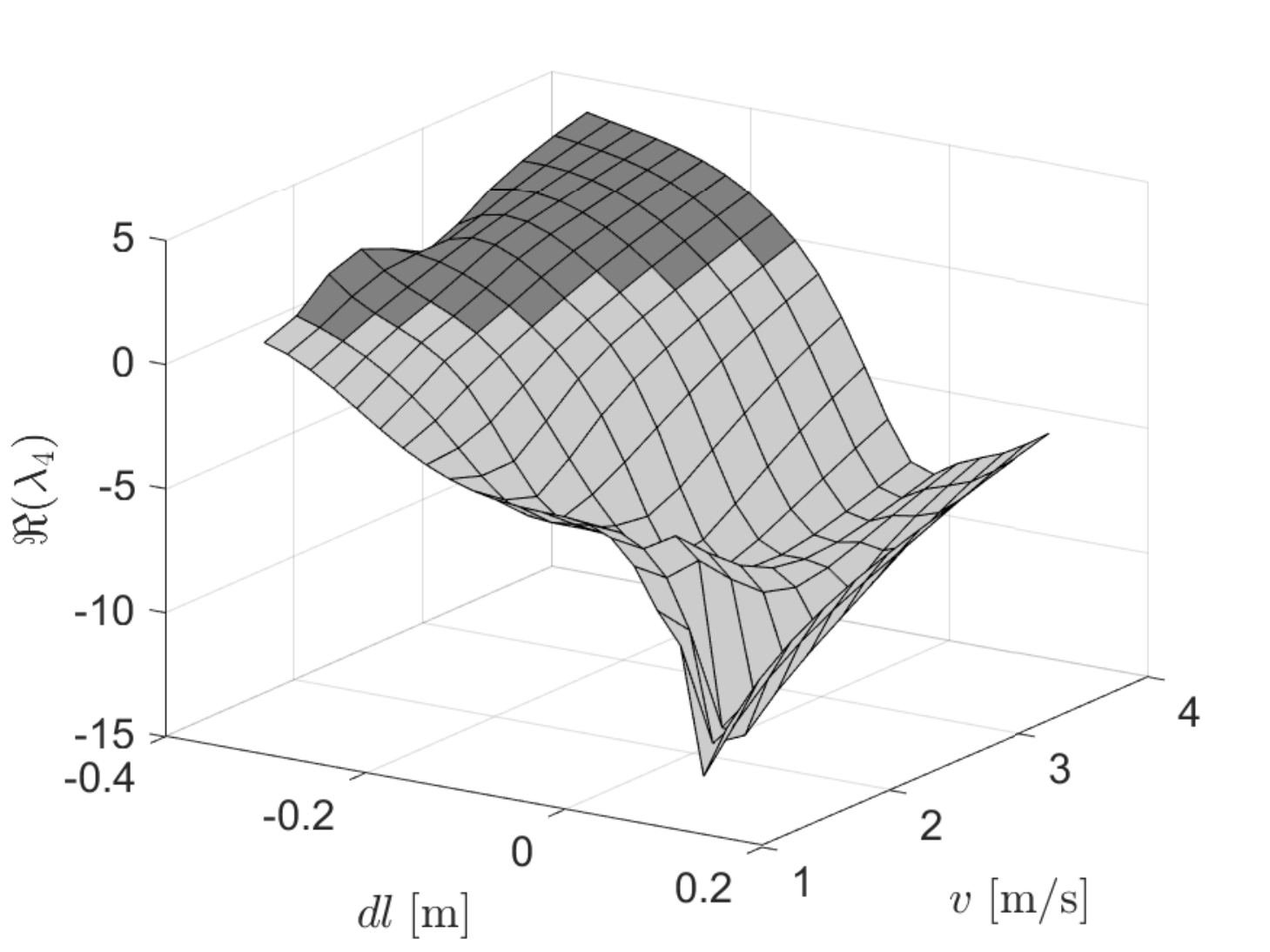}
	\caption{Real part of eigenvalue $\lambda_4$ of the dynamics of $\xi$, linearised around the defined equilibrium value. Light gray is used to define the region in which $\mathcal{R}(\lambda_4)<0$, while dark gray is used to define the region where $\mathcal{R}(\lambda_4)\geq 0$.}\label{fig:l4}
\end{figure}

Define $\xi=\begin{bmatrix}\psi&r&\beta&\delta\end{bmatrix}^T$. Its dynamics is governed by~\eqref{eq:singletrack_model},~\eqref{eq:deltadot}, and~\eqref{eq:feedback_lin_uncert}, and it is independent of ${x}_{{\rm\scriptscriptstyle P}}$ and ${y}_{{\rm\scriptscriptstyle P}}$. For this reason, the stability analysis described in this section focuses on $\xi$ only. This is necessary, but also sufficient, for unravelling the properties of the overall system dynamics.\\
Considering the described setup, $p$ is selected equal to $35\,\mathrm{cm}$, $v_{{\rm\scriptscriptstyle P},x}=\bar{v}\COS{\bar{\psi}}$ and $v_{{\rm\scriptscriptstyle P},y}=\bar{v}\SIN{\bar{\psi}}$. Therefore, the stability of the steady motion $\psi=\bar{\psi}$, $r=0$, $\beta=0$, and $\delta=0$ is analysed as parameter $dl$ varies in the range of its possible values for all reasonable values of $\bar{v}$. Without loss of generality, $\bar{\psi}$ is selected equal to $\pi/4$.\\
This analysis is conducted using MatCont software tool~\cite{bib:DhoogeEtAl2008} and its results are summarised in Figure~\ref{fig:bifurcaz}. Subcritical Hopf bifurcations~\cite{bib:DercoleRinaldi2011} are detected at negative values of $dl$, but these values are not physically meaningful, since they fall outside the range $[-l_f,l_r]$ for all considered velocity values. Therefore, the considered equilibrium point is structurally asymptotically stable.
For a more thorough insight, in Figures~\ref{fig:l1}-\ref{fig:l4} we show the values of the real parts of the four eigenvalues of the dynamics of $\xi$, linearised around the defined equilibrium value.
Consistently with Figure~\ref{fig:bifurcaz}, the latter figures show that, in physically meaningful ranges of parameter $dl$, $\mathcal{R}(\lambda_i)<0$ for all $i=1,\dots,4$.
%

\section{Experimental results} \label{sec:exp-results}
This section reports the results of an experimental campaign conducted on the vehicle described in Section~\ref{sec:exp-setup} to show the effectiveness of the proposed control approach. First, we show the results of the tests executed on the feedback linearised system only, to verify that it behaves as predicted by the theory, i.e., as two independent integrators, even in the presence of uncertainty. Secondly, we present trajectory tracking results, achieved with a simple tracking controller constituted by an outer proportional position loop and an inner feedback linearising law. All these results have been achieved selecting for point $\rm{P}$ a distance $p$ equal to $0.35\,\mathrm{m}$ from the front wheel contact point.
\begin{figure}[htbp]
	\begin{center}
      \includegraphics[width=0.475\columnwidth]{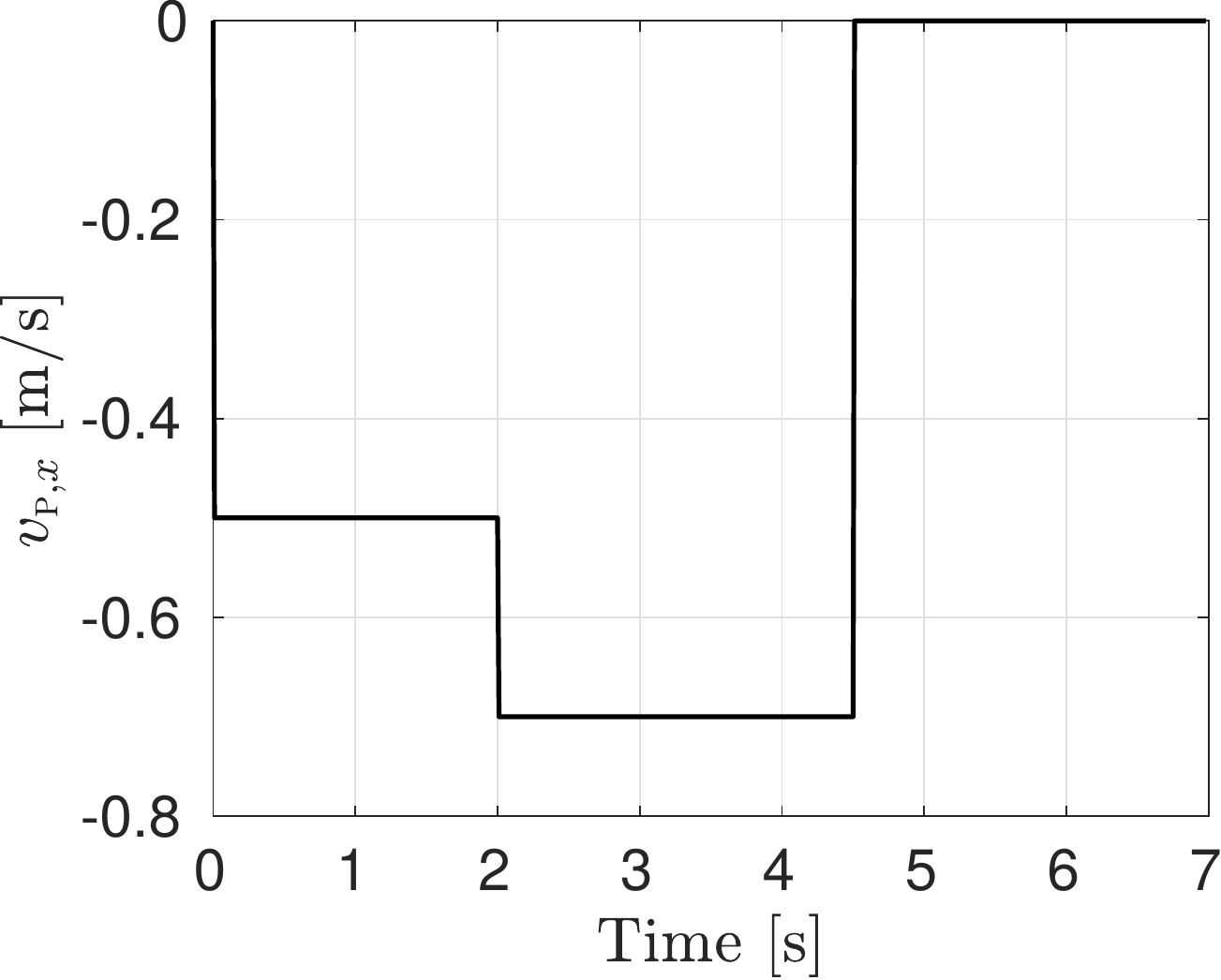}\hfill
      \includegraphics[width=0.49\columnwidth]{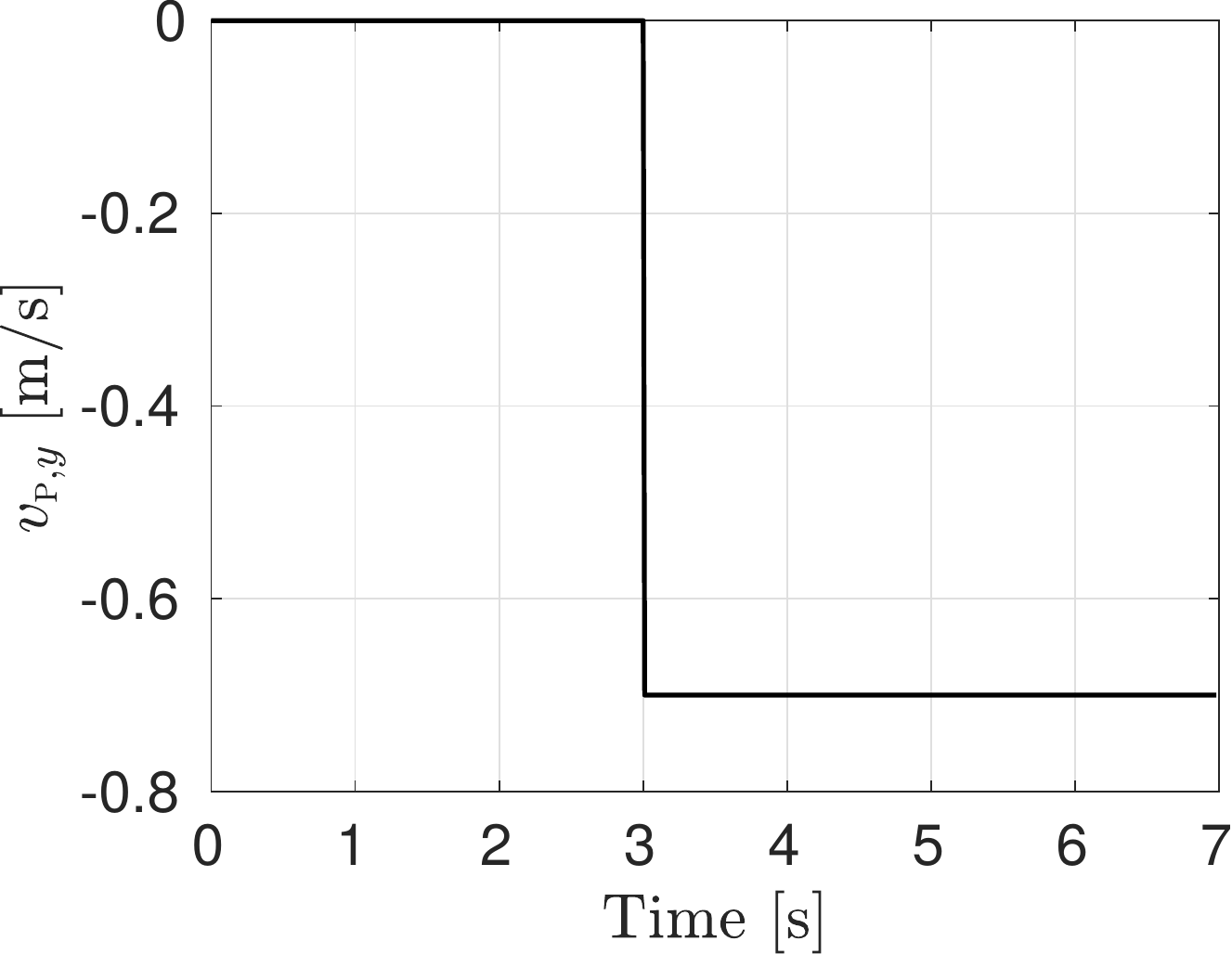}
      \caption{Commanded velocity of point $\rm{P}$.\label{fig:exp-ol-vp}}
    \end{center}
\end{figure}
\subsection{Feedback linearising law test}
A first set of open-loop tests has been performed in order to verify that, despite non-idealities introduced by vehicle and tyre-ground interaction, the closed-loop system behaves as predicted by the theory, i.e., as two independent integrators relating velocity along $x$ and $y$ axis to position of point $\rm{P}$.
\begin{figure}[htbp]
	\centering
    \subfloat[$dl=0\,\mathrm{m}$]{%
         \includegraphics[width=0.48\columnwidth]{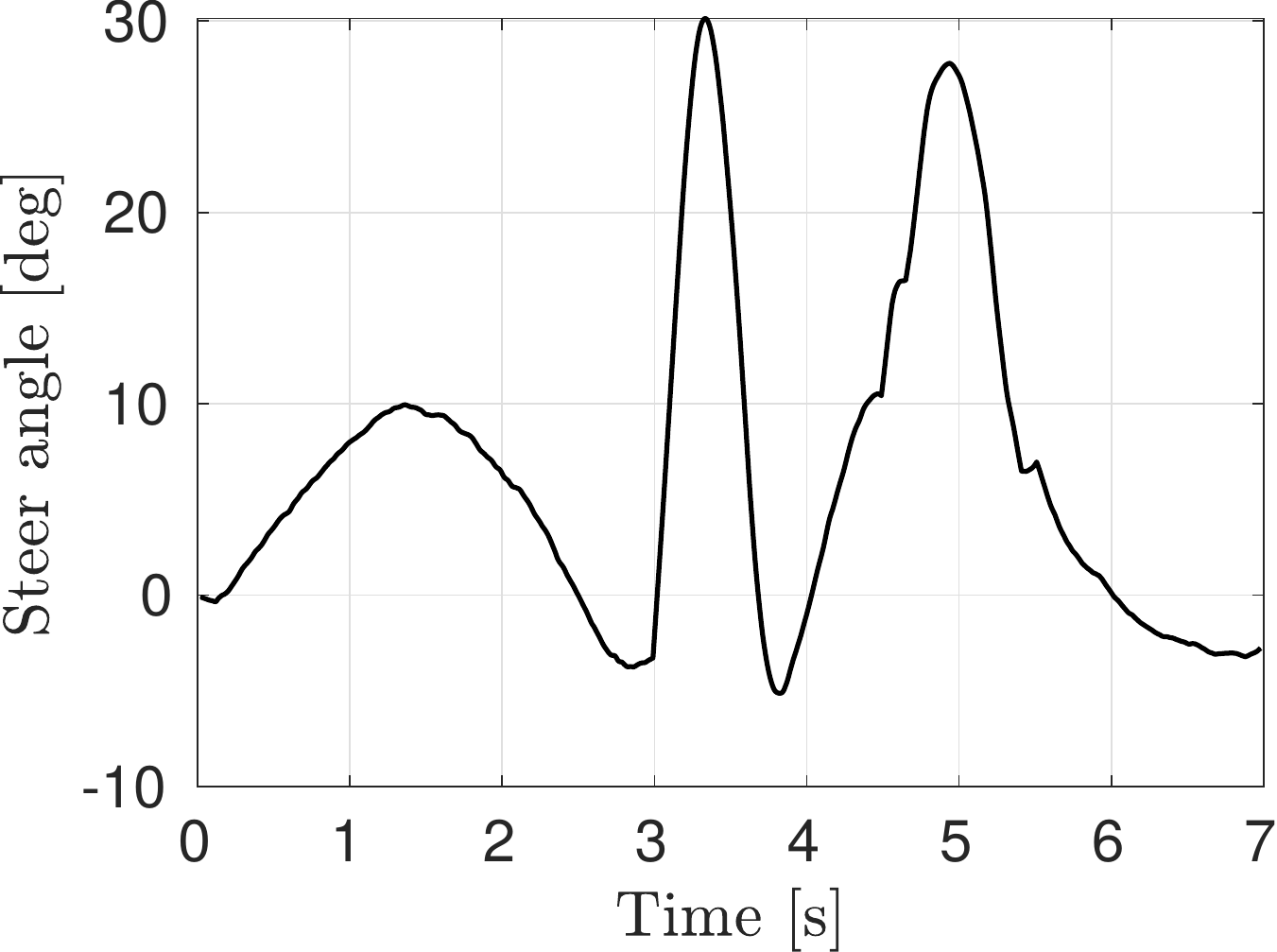}}
    \hfill
    \subfloat[$dl=0.132\,\mathrm{m}$]{%
         \includegraphics[width=0.48\columnwidth]{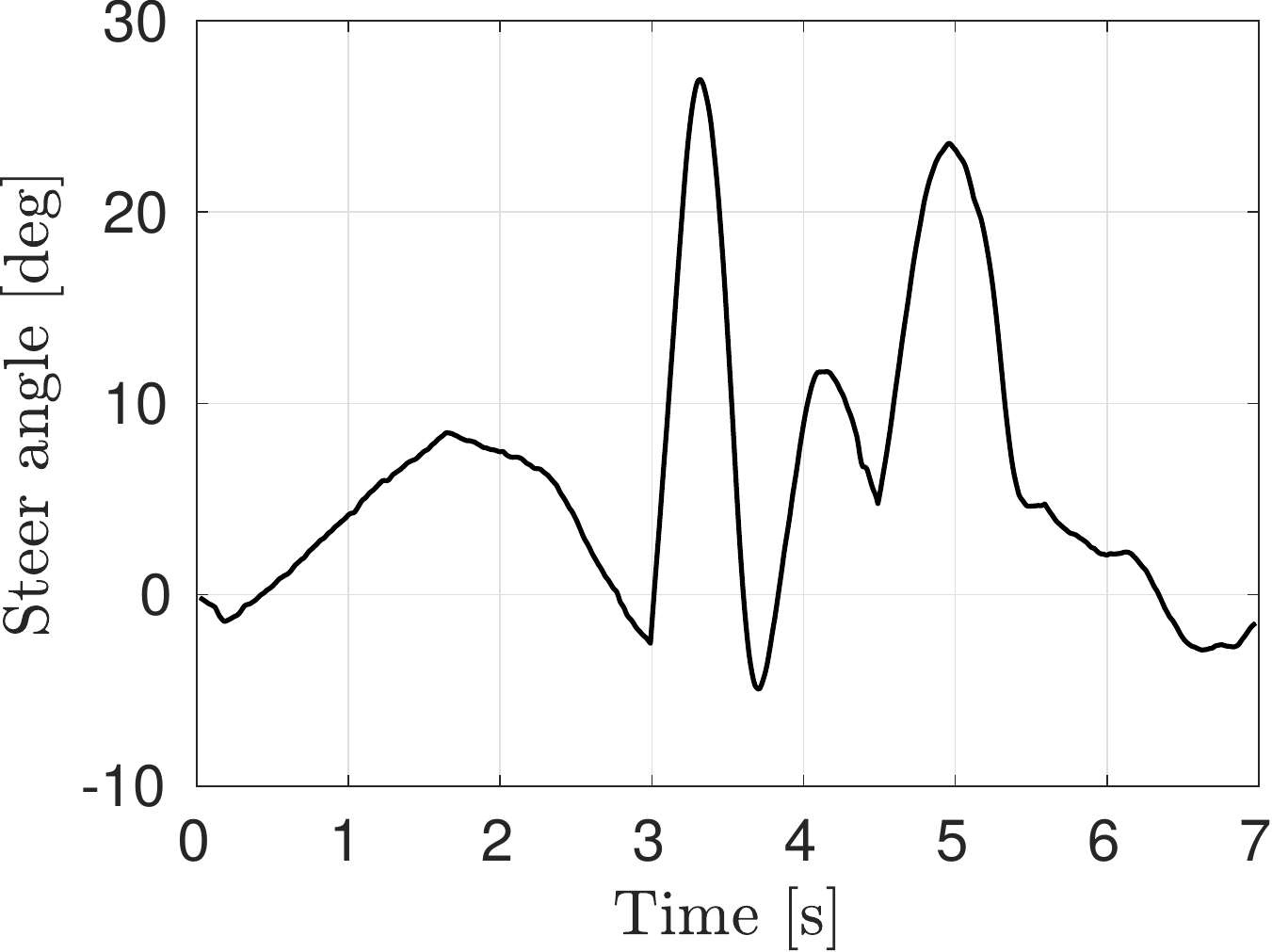}}
    \caption{Reference steer position generated by the feedback linearising law.\label{fig:exp-ol-steer}}
\end{figure}
This section reports the results of two of these experiments, a first one in which $l_f^{\rm\scriptscriptstyle est}$ has been set equal to $0.1368\,\mathrm{m}$, i.e., the value obtained by the identification procedure, and a second one in which $dl$ is equal to $0.1232\,\mathrm{m}$, as if the center of mass were located at the rear wheel contact point. For both tests, the feedback linearising law~\eqref{eq:feedback_lin_uncert} has been fed with the same piecewise constant velocity signals (Figure~\ref{fig:exp-ol-vp}), that allow to perform a sequence of step responses.
\begin{figure}[htbp]
	\centering
    \subfloat[$dl=0\,\mathrm{m}$]{%
         \includegraphics[width=0.48\columnwidth]{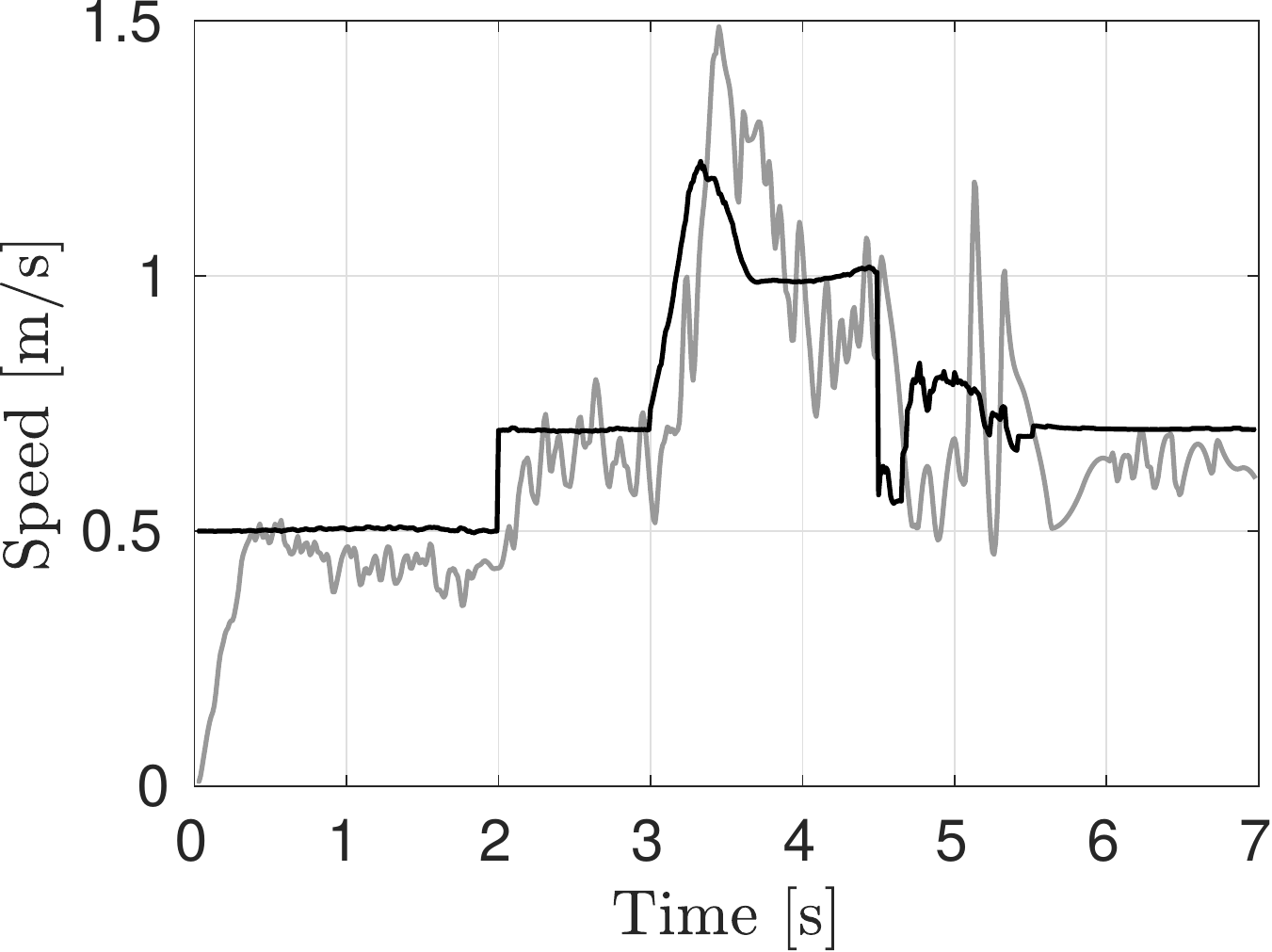}}
    \hfill
    \subfloat[$dl=0.132\,\mathrm{m}$]{%
         \includegraphics[width=0.48\columnwidth]{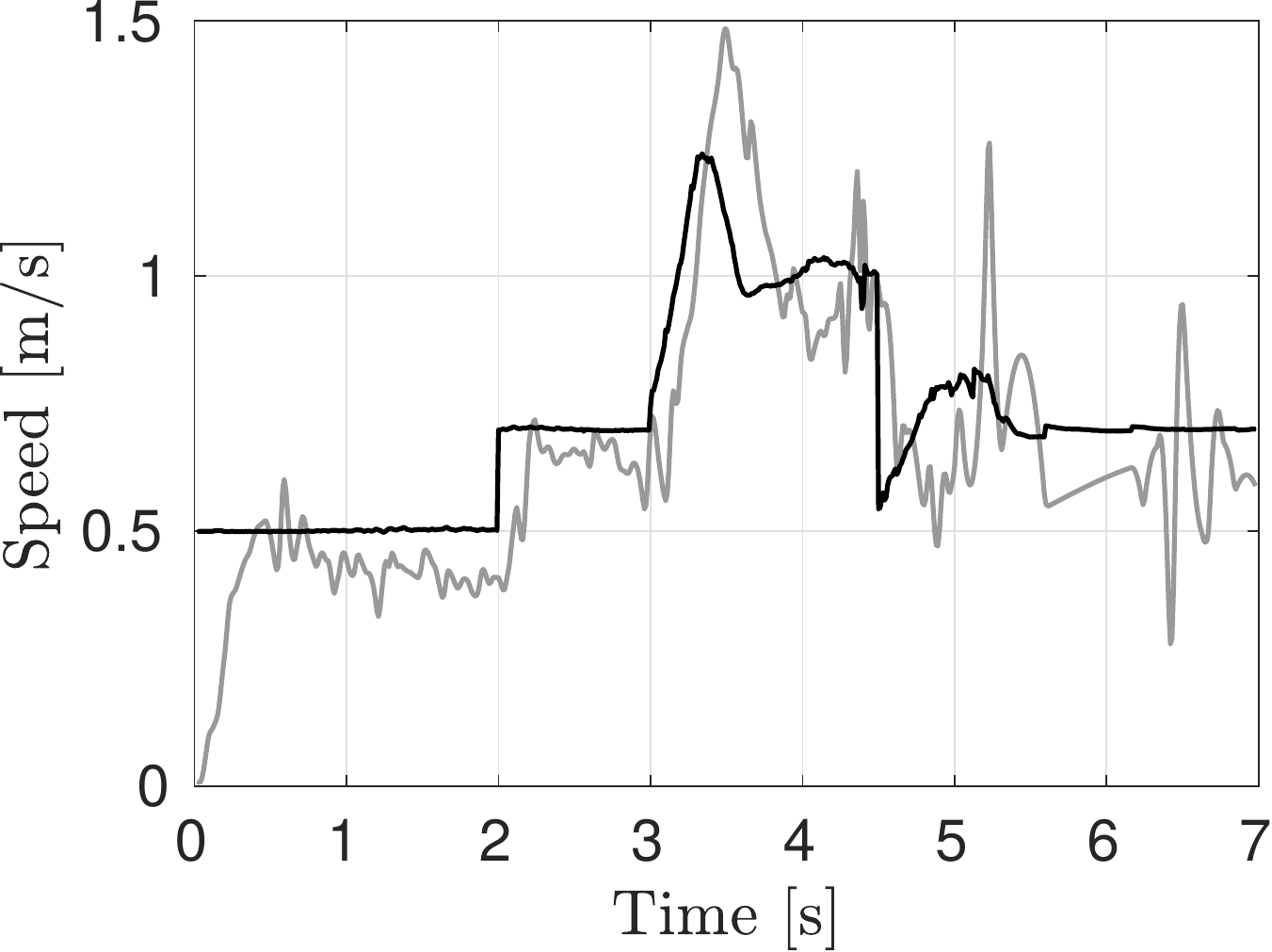}}
    \caption{Reference speed generated by the feedback linearising law (gray line) and actual vehicle speed (black line).\label{fig:exp-ol-speed}}
\end{figure}
\begin{figure}[htbp]
	\centering
    \subfloat[$dl=0\,\mathrm{m}$]{%
         \includegraphics[width=0.49\columnwidth]{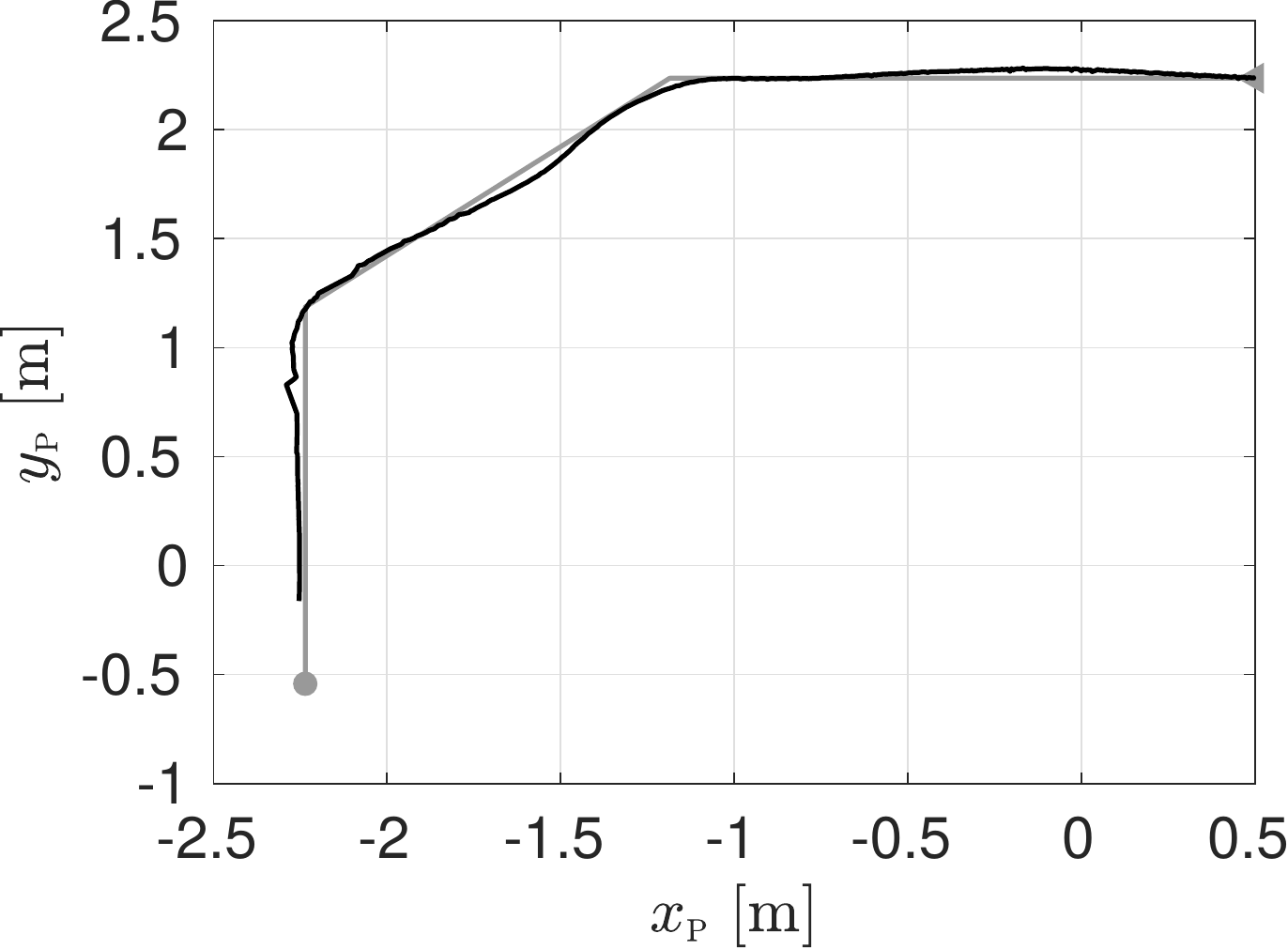}}
    \hfill
    \subfloat[$dl=0.132\,\mathrm{m}$]{%
         \includegraphics[width=0.47\columnwidth]{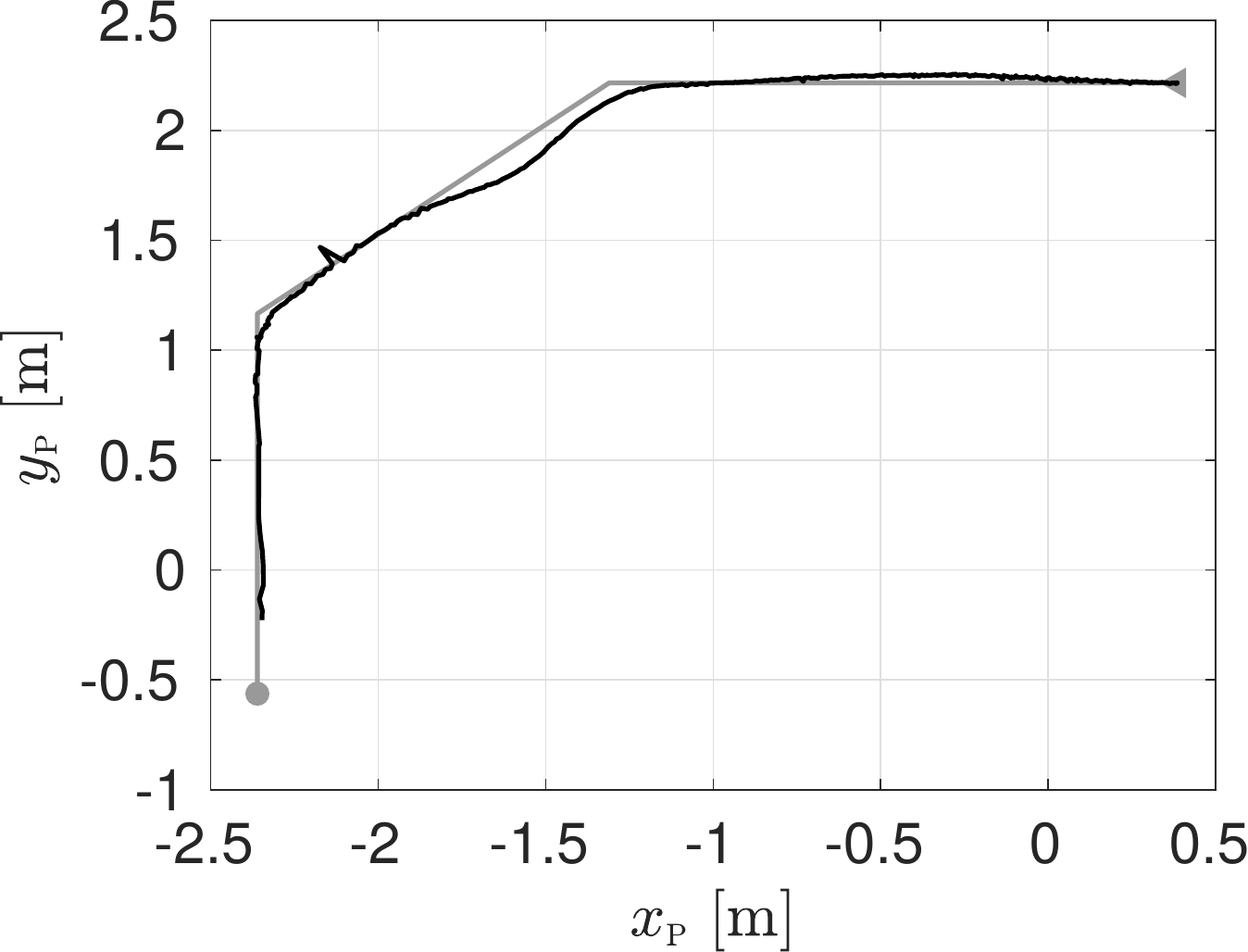}}
      \caption{Nominal (gray line) and actual (black line) trajectory of point $\rm{P}$. Each trajectory starts from the triangle and ends in the circle.\label{fig:exp-ol-trajectory}}
\end{figure}

\begin{figure}[htbp]
	\centering
	\subfloat[$dl=0\,\mathrm{m}$]{%
         \includegraphics[width=0.475\columnwidth]{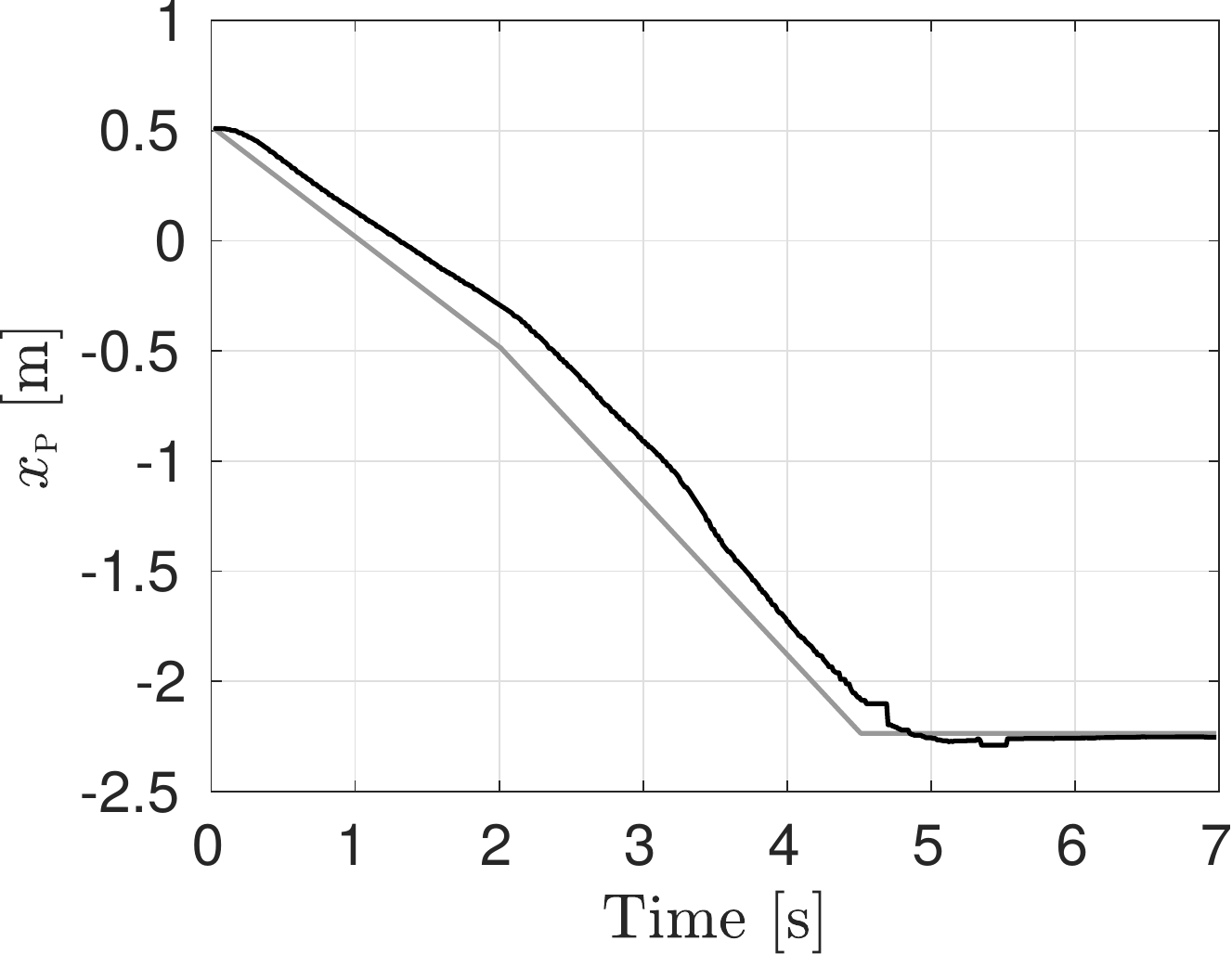}\hspace{5pt}
         \includegraphics[width=0.49\columnwidth]{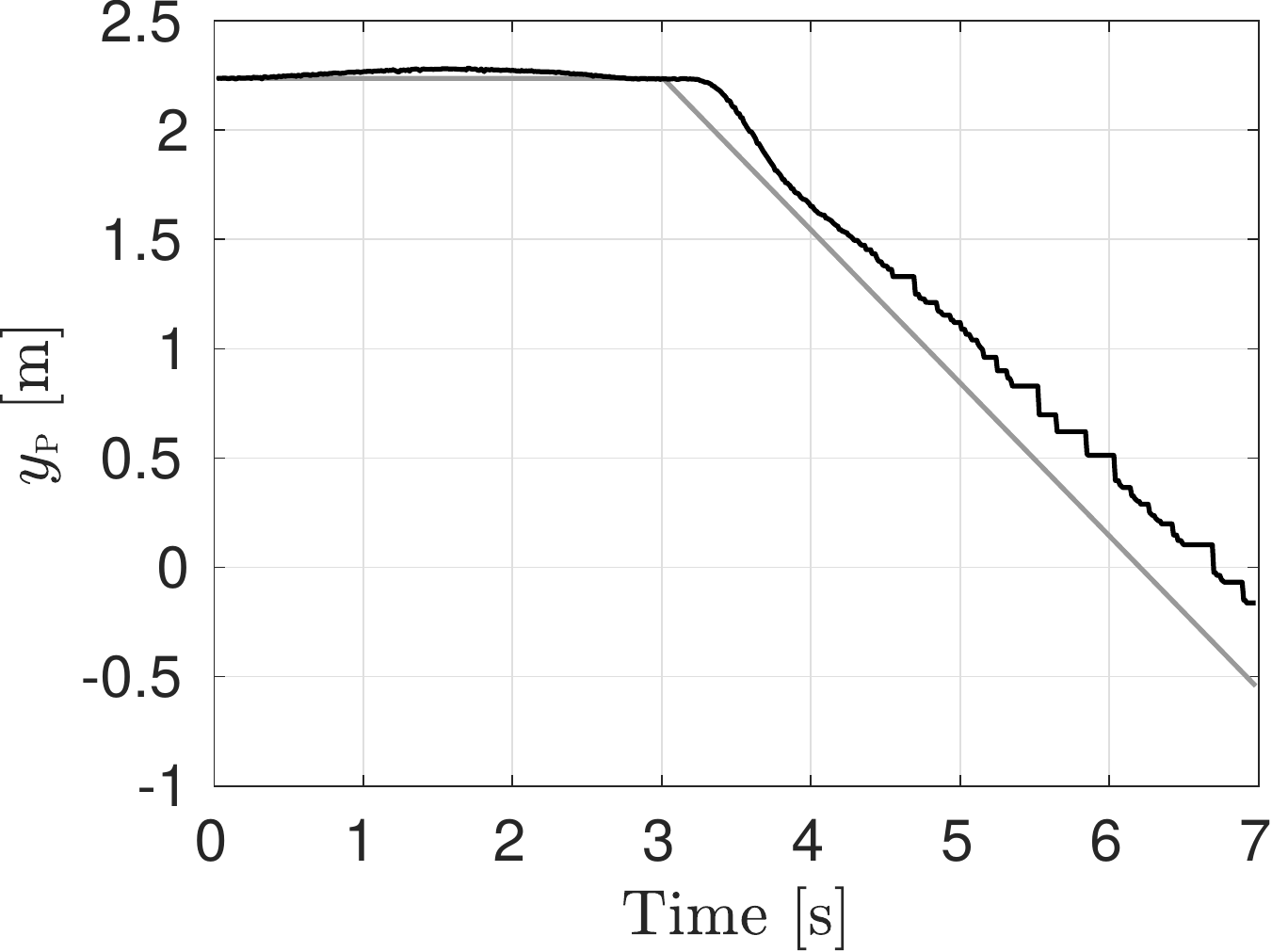}}\\
	\subfloat[$dl=0.132\,\mathrm{m}$]{%
         \includegraphics[width=0.48\columnwidth]{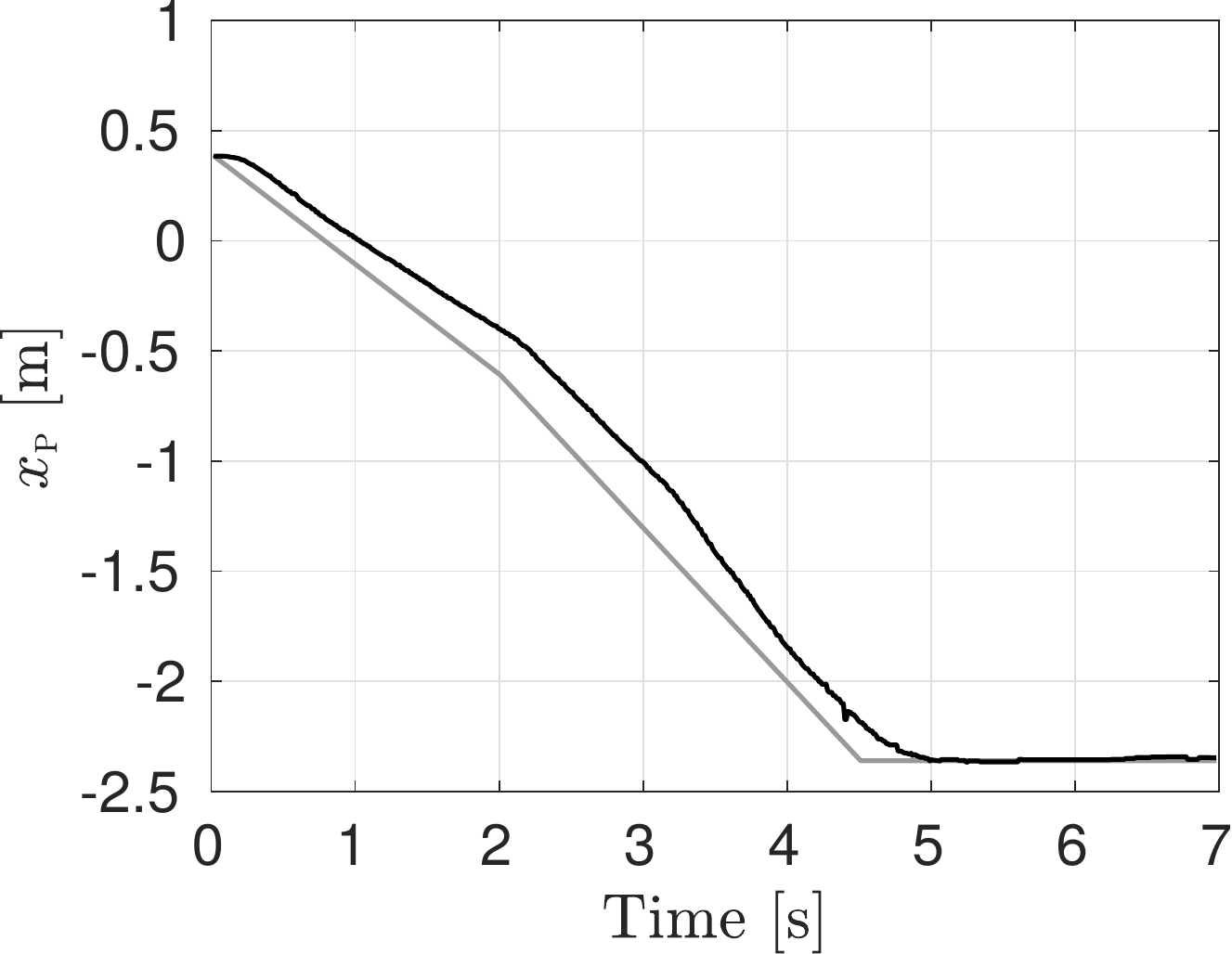}\hspace{5pt}
         \includegraphics[width=0.48\columnwidth]{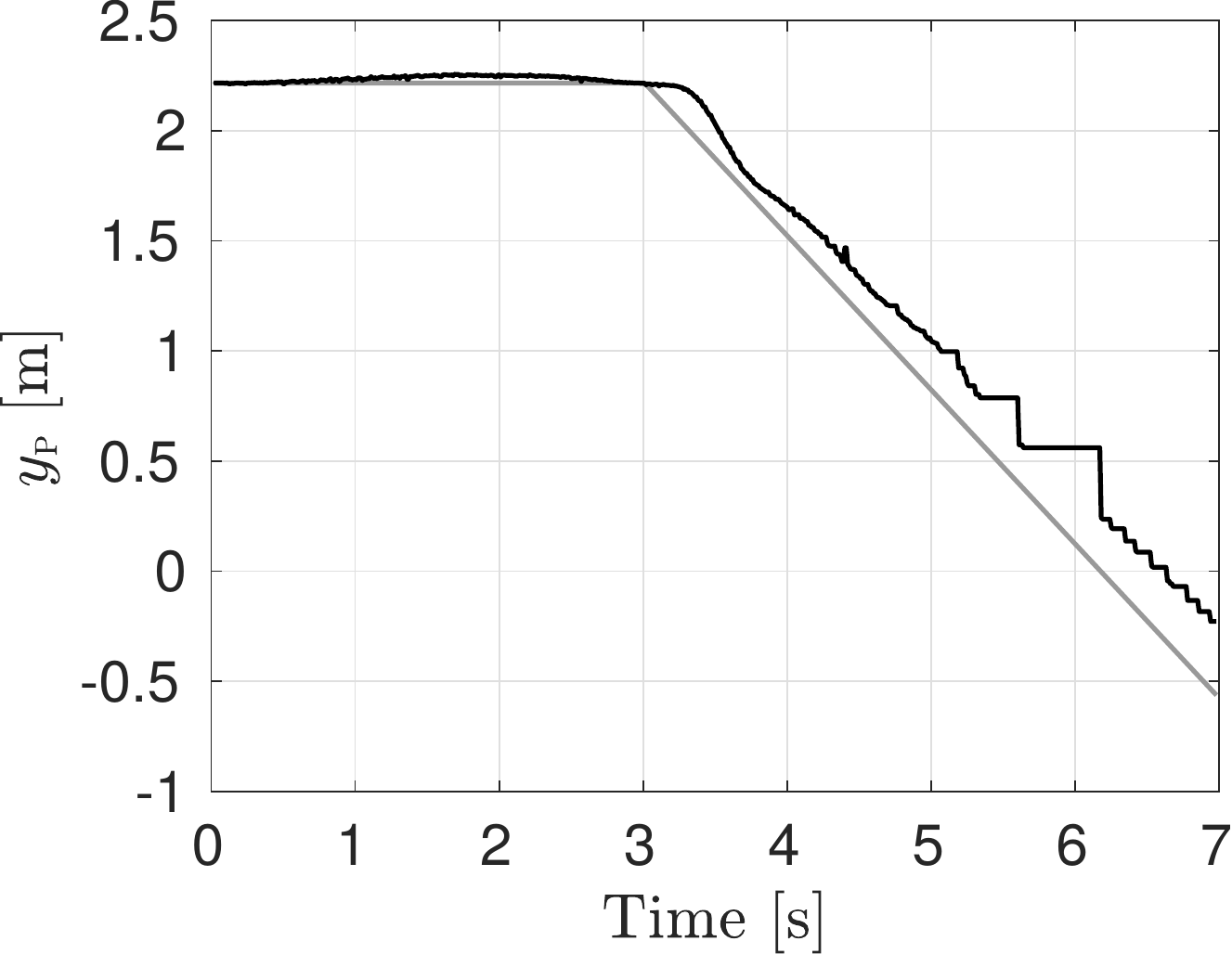}}
    \caption{Nominal (gray line) and actual (black line) position of point $\rm{P}$.\label{fig:exp-ol-xyp}}
\end{figure}

Figures~\ref{fig:exp-ol-trajectory} and~\ref{fig:exp-ol-xyp} report the actual (black line) and the nominal trajectory (gray line) of point $\rm{P}$, obtained integrating system~\eqref{eq:double_integratorP}. As it can be seen from these figures, though there is no position control, despite non-idealities and disturbances, and even in the case of a significant error in the center of mass estimation (note that $0.1232\,\mathrm{m}$ is the upper bound of the range of physically meaningful values), the trajectory performed by the vehicle is very close to the nominal one, showing the remarkable performance and robustness of the proposed feedback linearising law.

Note that the markers used by the optical system to track vehicle position are occasionally lost by the cameras, due to occlusions. As a consequence, no pose measurement is generated and the controller is forced to keep the last available heading and position measurements. This can be clearly seen in the time evolution of $x$ and $y$ positions (Figure~\ref{fig:exp-ol-xyp}), that keep sometime constant even if the vehicle is moving. The results, however, show that the linearising feedback is robust to this disturbance as well.
\begin{figure*}[htbp]
\tikzstyle{KPx} = [draw, rectangle, minimum height=5.25mm, minimum
width=5mm,align=center]
\tikzstyle{KPy} = [draw, rectangle, minimum height=5.25mm, minimum
width=5mm,align=center]
\tikzstyle{controller} = [draw, rectangle, minimum height=15mm, minimum
width=14mm,align=center]
\tikzstyle{model} = [draw, rectangle, minimum height=20mm, minimum
width=12mm,align=center,line width=0.5mm,fill=lightgray]
\tikzstyle{integrator} = [draw, rectangle, minimum height=5.25mm, minimum
width=5mm,align=center]
\tikzstyle{Ptransform} = [draw, rectangle, minimum height=12mm, minimum
width=8mm,align=center]
\centering
\begin{tikzpicture}
 \node[controller] (a) at (-0.52,0) {\scriptsize Feedback\\[-0.1cm] \scriptsize
linearising\\[-0.1cm] \scriptsize law};
 \node[integrator] (c) at (1.0,-0.5) {$\int$};
 \node[model] (b) at (2.42,0) {\scriptsize Vehicle};
 \node[Ptransform] (d) at (4.74,0.4) {\scriptsize Point $\rm{P}$\\[-0.1cm] \scriptsize transform};
 \node[KPx] (e) at (-2.55,0.47) {$K_{P_x}$};
 \node[KPy] (f) at (-2.55,-0.45) {$K_{P_y}$};
 \draw[->] (-2.1,0.5) -- (-1.22,0.5) node[above,pos=0.45] {$v_{{\rm\scriptscriptstyle P},x}$};
 \draw[->] (-2.1,-0.5) -- (-1.22,-0.5) node[above,pos=0.45] {$v_{{\rm\scriptscriptstyle P},y}$};
 \draw[->] (0.18,0.5) -- (1.8,0.5) node[above,pos=0.5] {$v$};
 \draw[->] (0.18,-0.5) -- (0.75,-0.5) node[above,pos=0.5] {$u_{\delta}$};
 \draw[->] (1.25,-0.5) -- (1.8,-0.5) node[above,pos=0.5] {$\delta$};
 \draw[->] (3.04,-0.8) -- (4.14,-0.8) node[above,pos=0.5] {$\psi,\beta,\delta$};
 \draw[->] (3.04,0.75) -- (4.14,0.75) node[above,pos=0.5] {$x_{{\rm\scriptscriptstyle G}},y_{{\rm\scriptscriptstyle G}}$};
 \draw[->] (3.04,0) -- (4.14,0) node[above,pos=0.5] {$\psi,\delta$};
 \draw[->] (5.34,0) -- (6.1,0) node[above,pos=0.6] {$y_{{\rm\scriptscriptstyle P}}$};
 \draw[->] (5.34,0.75) -- (6.1,0.75) node[above,pos=0.6] {$x_{{\rm\scriptscriptstyle P}}$};
 \draw[-] (3.6,-0.8) -- (3.6,-1.5);
 \draw[-] (3.6,-1.5) -- (-0.5,-1.5);
 \draw[->] (-0.5,-1.5) -- (-0.5,-0.75);
 \draw[-] (5.5,0.0) -- (5.5,-2.0);
 \draw[-] (5.5,-2.0) -- (-2.55,-2.0);
 \draw[->] (-2.55,-2.0) -- (-2.55,-0.75);
 \draw[-] (5.5,0.75) -- (5.5,1.6);
 \draw[-] (5.5,1.6) -- (-2.55,1.6);
 \draw[->] (-2.55,1.6) -- (-2.55,0.76);
 \draw[->] (-3.9,0.5) -- (-3.0,0.5) node[above,pos=0.5] {$x_{\rm\scriptscriptstyle P}^{\rm\scriptscriptstyle ref}$};
 \draw[->] (-3.9,-0.5) -- (-3.0,-0.5) node[above,pos=0.5] {$y_{\rm\scriptscriptstyle P}^{\rm\scriptscriptstyle ref}$};
\end{tikzpicture}
\caption{Trajectory tracking controller architecture.\label{fig:trajectory_tracking_scheme}}
\end{figure*}
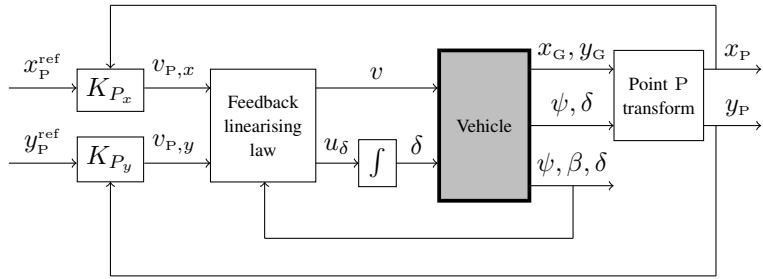
\begin{figure}[htbp]
	\begin{center}
      \includegraphics[width=0.7\columnwidth]{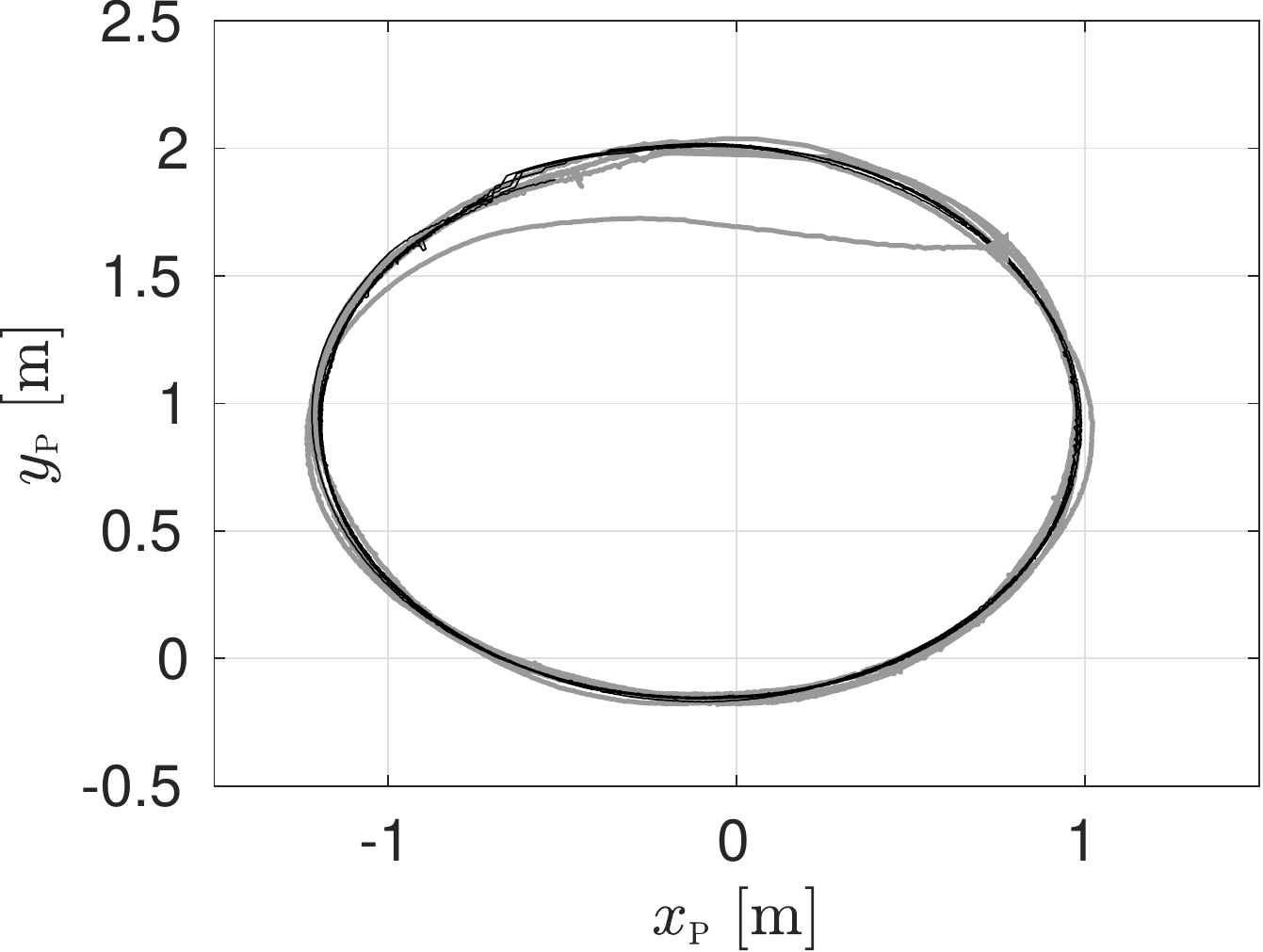}
      \caption{Reference (black line) and actual (gray line) trajectory of point $\rm{P}$. A triangle shows the starting position.\label{fig:exp-cl-circle-trajectory}}
    \end{center}
\end{figure}

\subsection{Trajectory tracking test}
The tests illustrated in this section aim at showing how the proposed feedback linearising law behaves when an outer trajectory tracking controller is considered. Though, as already mentioned in Section~\ref{sec:introduction}, any linear system tool can be applied in the design of the trajectory tracking controller, a simple proportional control has been used, as considering more complex strategies, like for example {MPC}, is out of the scope of this work.
\begin{figure}[htbp]
	\centering
    \includegraphics[width=0.475\columnwidth]{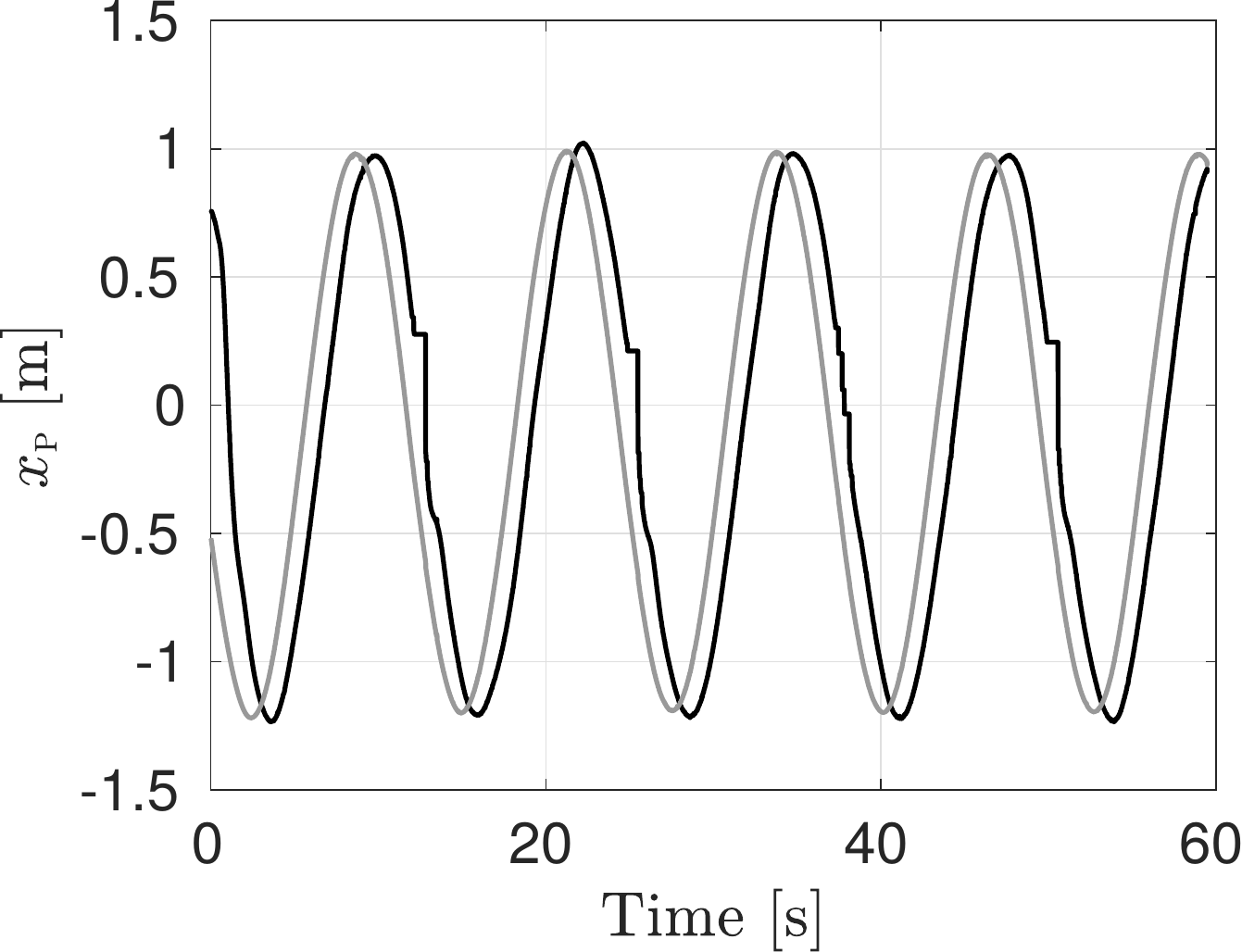} \hfill
    \includegraphics[width=0.49\columnwidth]{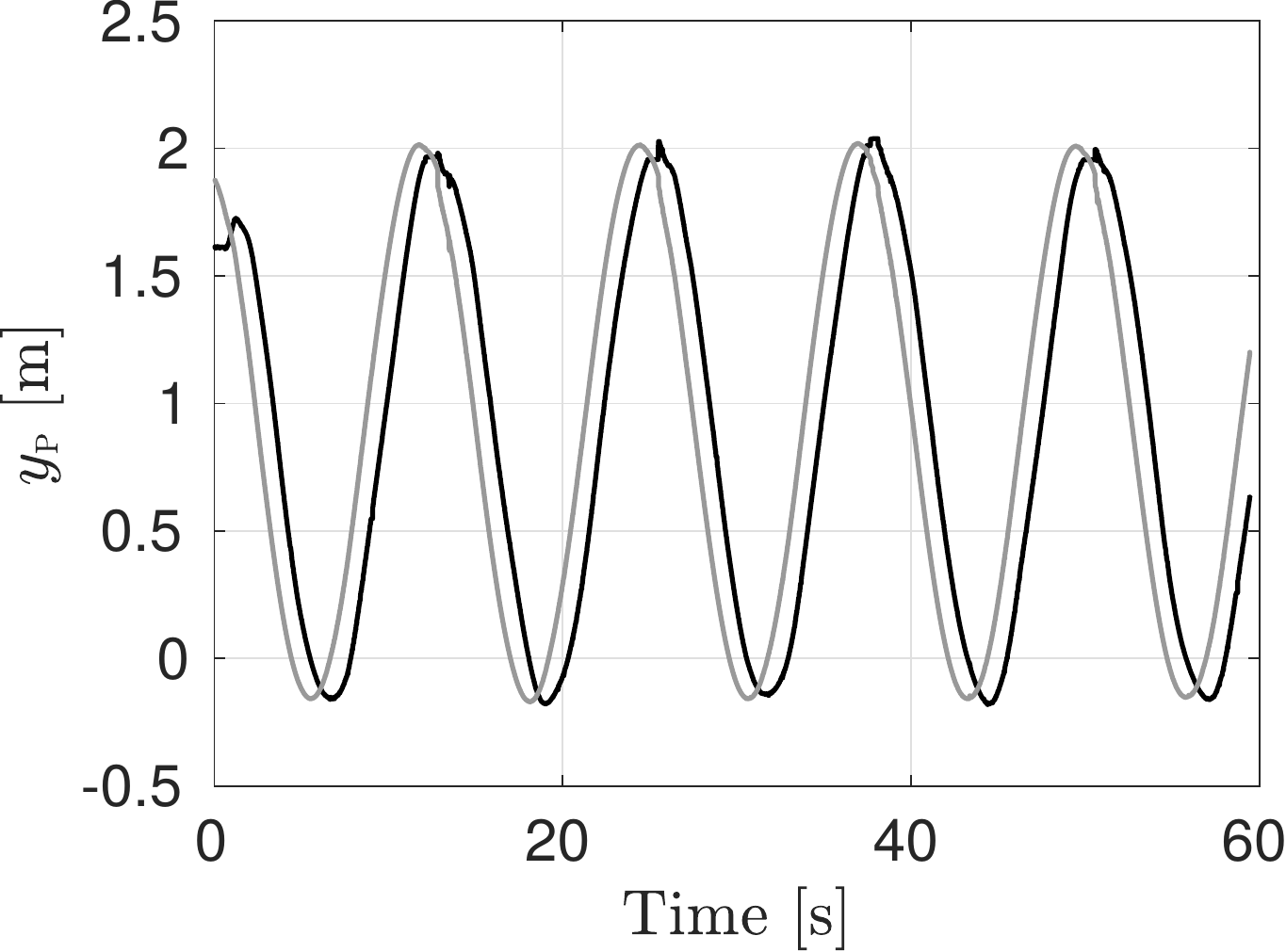}
    \caption{Reference (gray line) and actual (black line) position of point $\rm{P}$.\label{fig:exp-circle-xyp}}
\end{figure}

\begin{figure}[htbp]
	\centering
    \includegraphics[width=0.49\columnwidth]{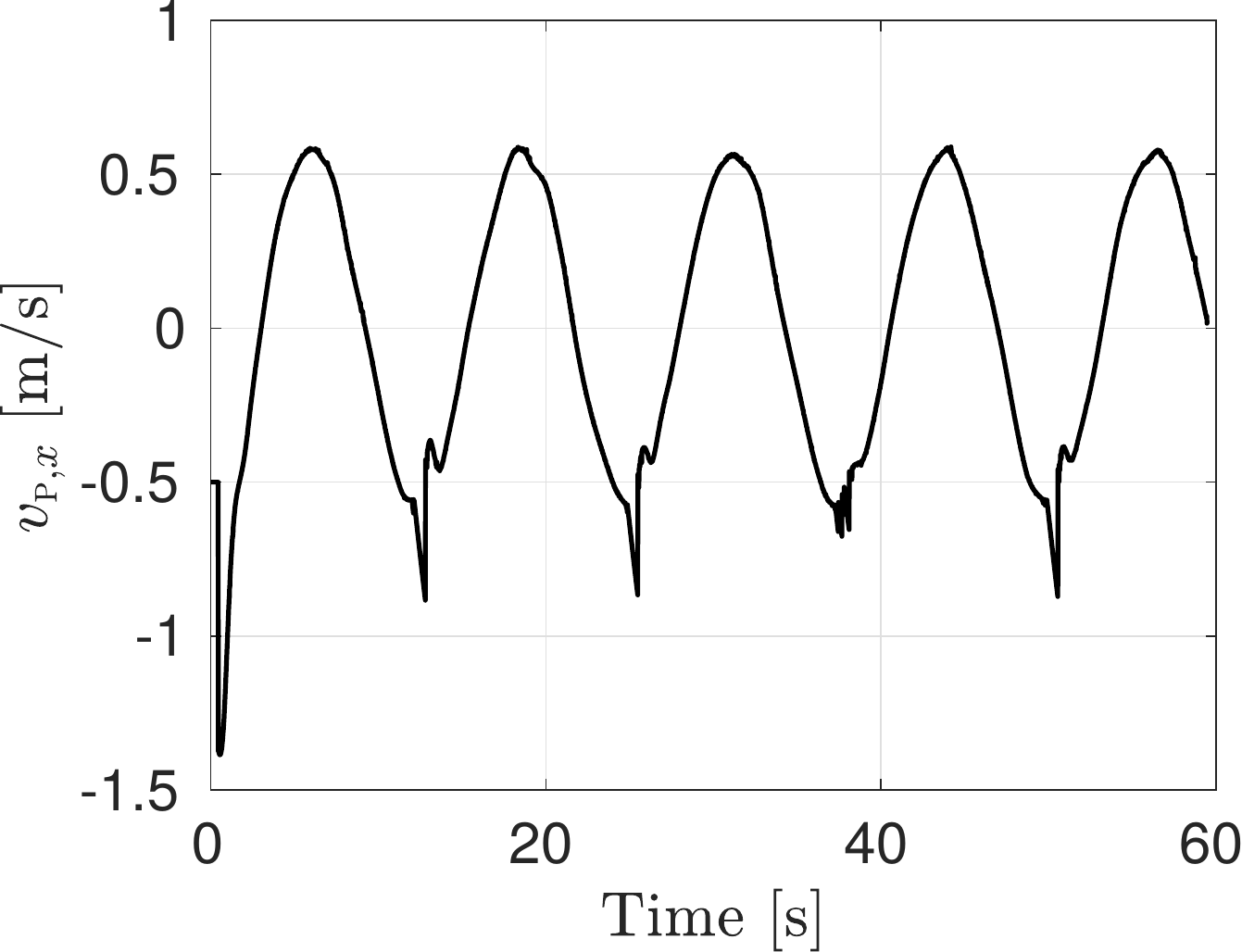} \hfill
    \includegraphics[width=0.47\columnwidth]{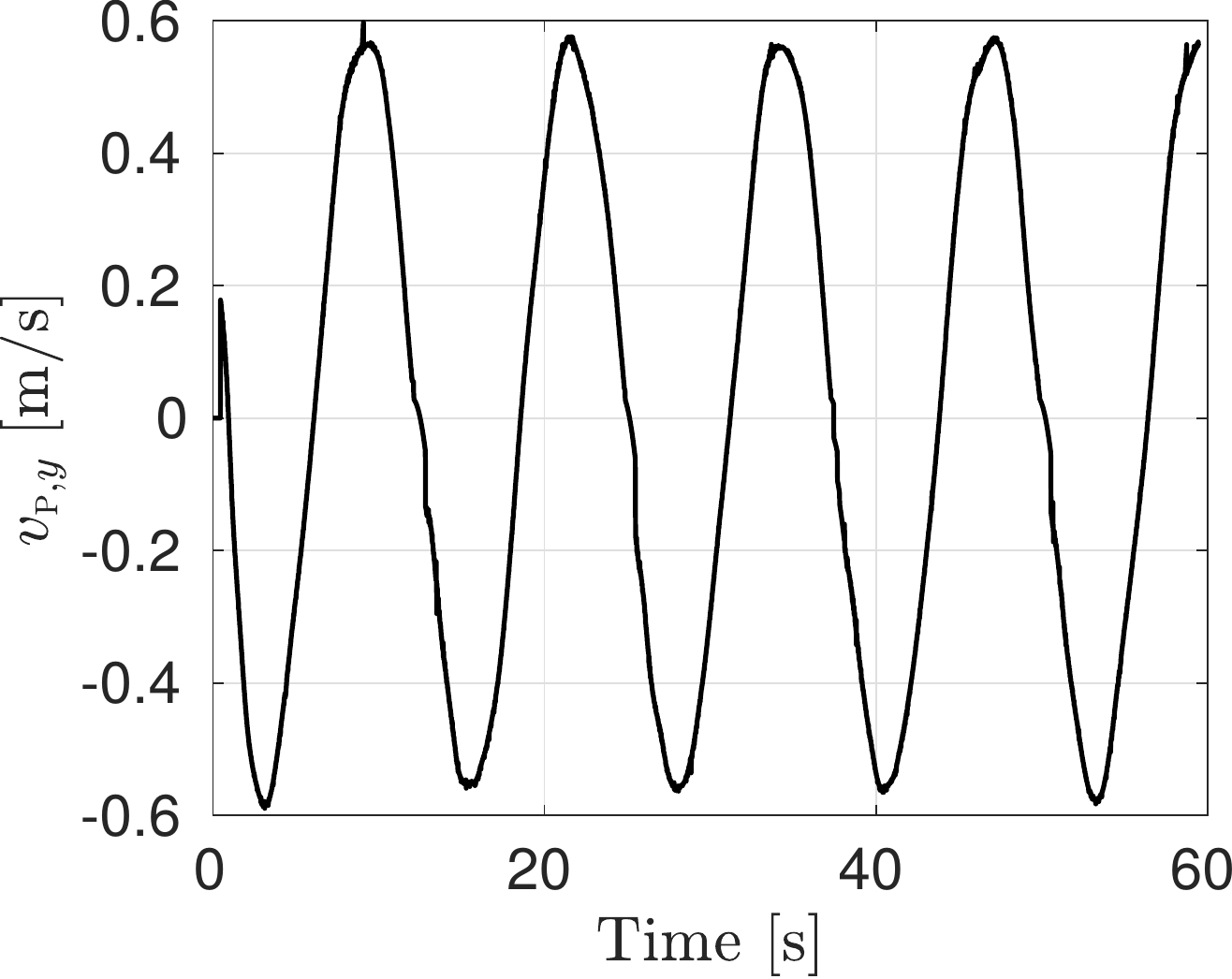}
    \caption{Velocity of point $\rm{P}$ generated by the position regulators.\label{fig:exp-circle-vp}}
\end{figure}

\begin{figure}[htbp]
   \centering
   \includegraphics[width=0.48\columnwidth]{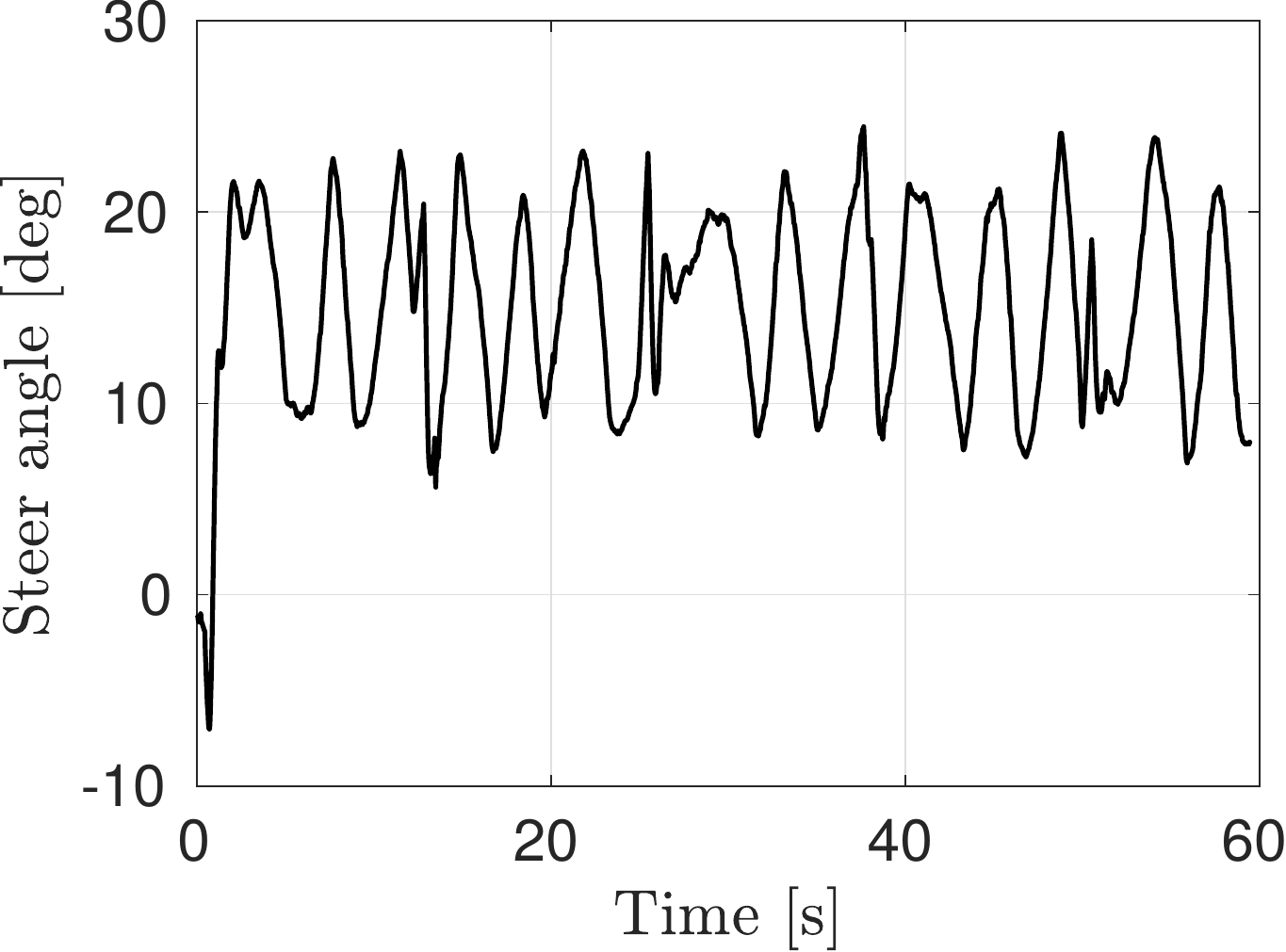} \hfill
   \includegraphics[width=0.48\columnwidth]{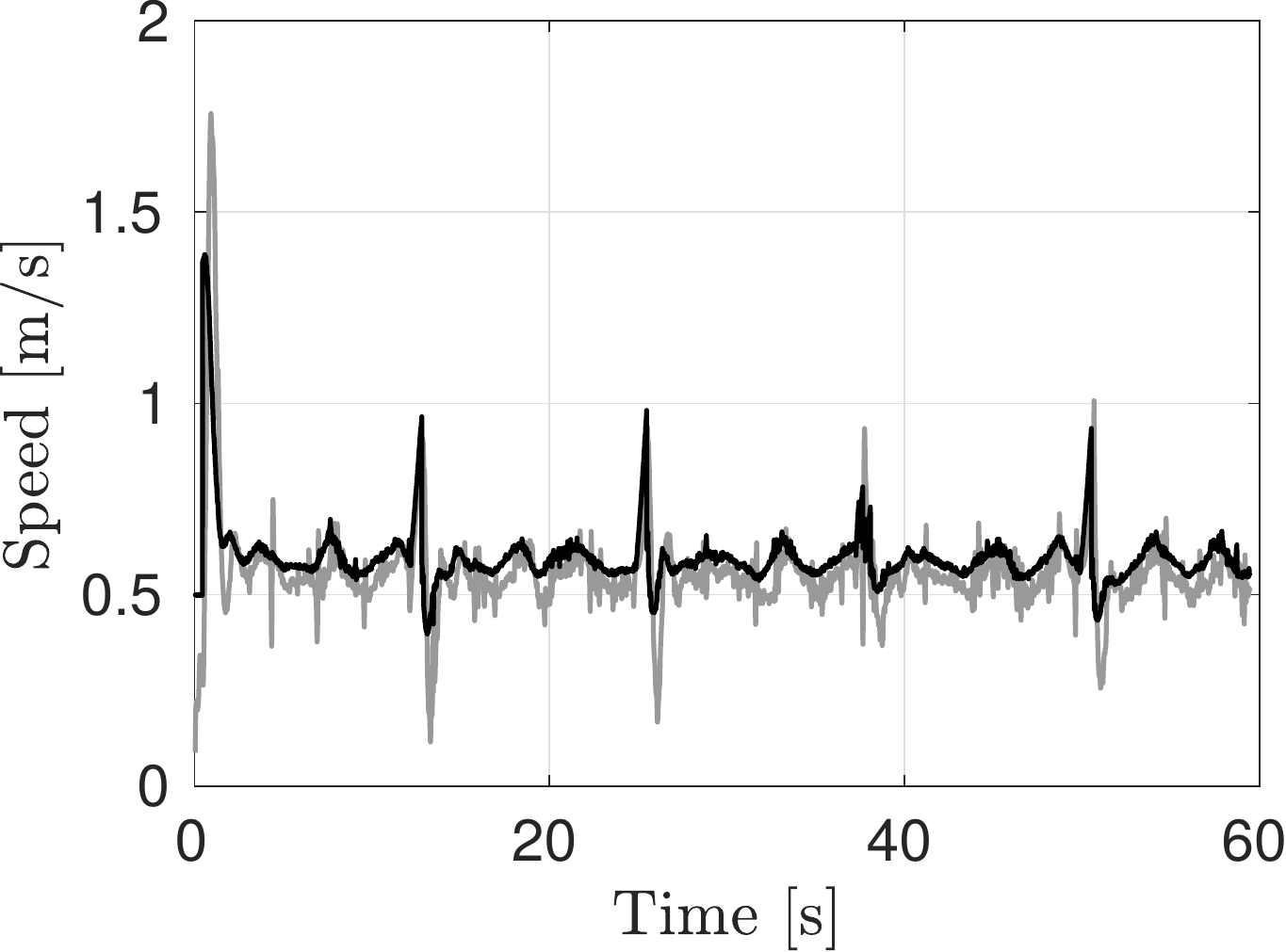}
   \caption{Reference steer position and reference speed (gray line) generated by the feedback linearising law, and actual vehicle speed (black line).\label{fig:exp-circle-cmd}}
\end{figure}

\begin{figure}[htbp]
   \centering
   \includegraphics[width=0.48\columnwidth]{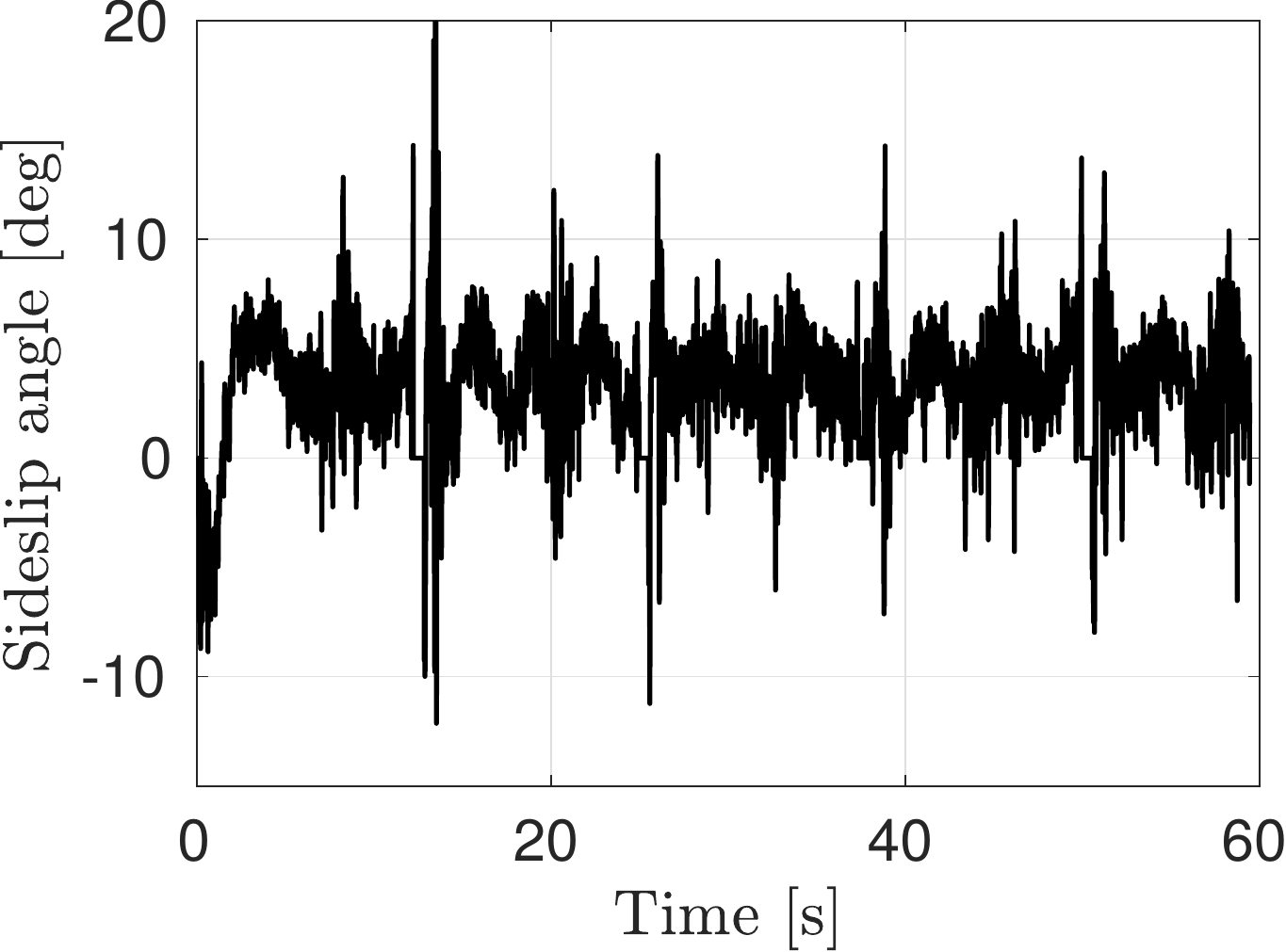} \hfill
   \includegraphics[width=0.48\columnwidth]{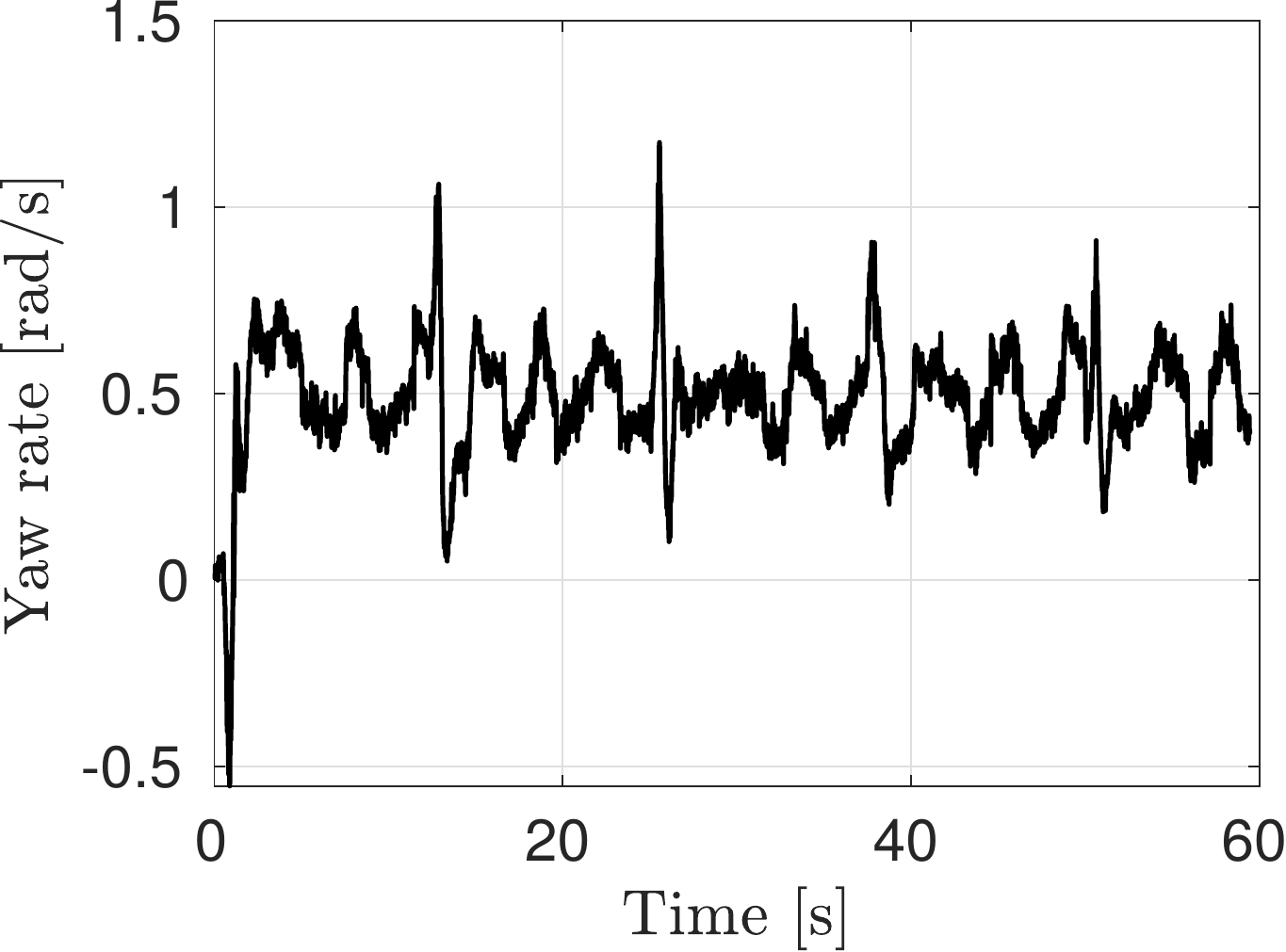}
   \caption{Sideslip angle and yaw rate.\label{fig:exp-circle-betar}}
\end{figure}

The same unitary gain has been selected for the two $x$ and $y$ proportional controllers $K_{P_x}$ and $K_{P_y}$ (Figure~\ref{fig:trajectory_tracking_scheme}), and a circular reference trajectory with a radius of $1\,\mathrm{m}$ and an angular velocity of $0.5\,\mathrm{rad/s}$ has been considered.
Figures~\ref{fig:exp-cl-circle-trajectory} and~\ref{fig:exp-circle-xyp} show the reference (gray line) and actual (black line) trajectory of point $\rm{P}$, showing the tracking performance that can be achieved with this very simple trajectory tracking controller. Note that, the reference and actual $x_{\rm\scriptscriptstyle P}$ and $y_{\rm\scriptscriptstyle P}$ positions are sinusoids with the same amplitude, demonstrating the vehicle is correctly tracking the reference path. On the other side, Figure~\ref{fig:exp-circle-xyp} shows a delay between the reference and actual positions, as the vehicle starts outside the desired path. Though this demonstrates the control system guarantees the convergence to the desired trajectory even in this situation, it causes, as a drawback, the time shift in the actual and reference positions shown in Figure~\ref{fig:exp-circle-xyp}. Figures~\ref{fig:exp-circle-vp},~\ref{fig:exp-circle-cmd}, and~\ref{fig:exp-circle-betar} report the time behaviour of the trajectory tracking control variables,  the outputs of the linearising law, and the measured sideslip and yaw rate, respectively.

These results demonstrate that the proposed feedback linearising law represents a tool to support the development of a generic, simple, and flexible control methodology for mobile robots affected by nonholonomic constraints.

\section{Conclusions} \label{sec:conclusion}
A simple and robust algorithm, that allows to feedback linearise a single-track dynamic model, has been presented in this paper. Differently from the classical feedback linearising law developed for mechanical systems modelled using the Lagrangian formulation, that depends on possibly uncertain dynamic parameters and suffers from robustness issues, the one here proposed depends only on the center of mass position. A numerical bifurcation analysis demonstrates that, for physically meaningful deviations of this parameter from its nominal value, the equilibrium point is structurally asymptotically stable. Finally, experimental results confirm the robustness of the linearising law, and show how using an inner linearising loop can simplify the design of a trajectory tracking controller, achieving satisfactory tracking performance even with a proportional position control law.

\section*{Appendix}
This section introduces an alternative feedback linearising law for a single-track dynamic model, different from the one presented in Section~\ref{sub:sec:feedback_linearization}, that is considered for the sake of comparison.
\begin{figure}[htbp]
	\begin{center}
      \includegraphics[width=0.8\columnwidth]{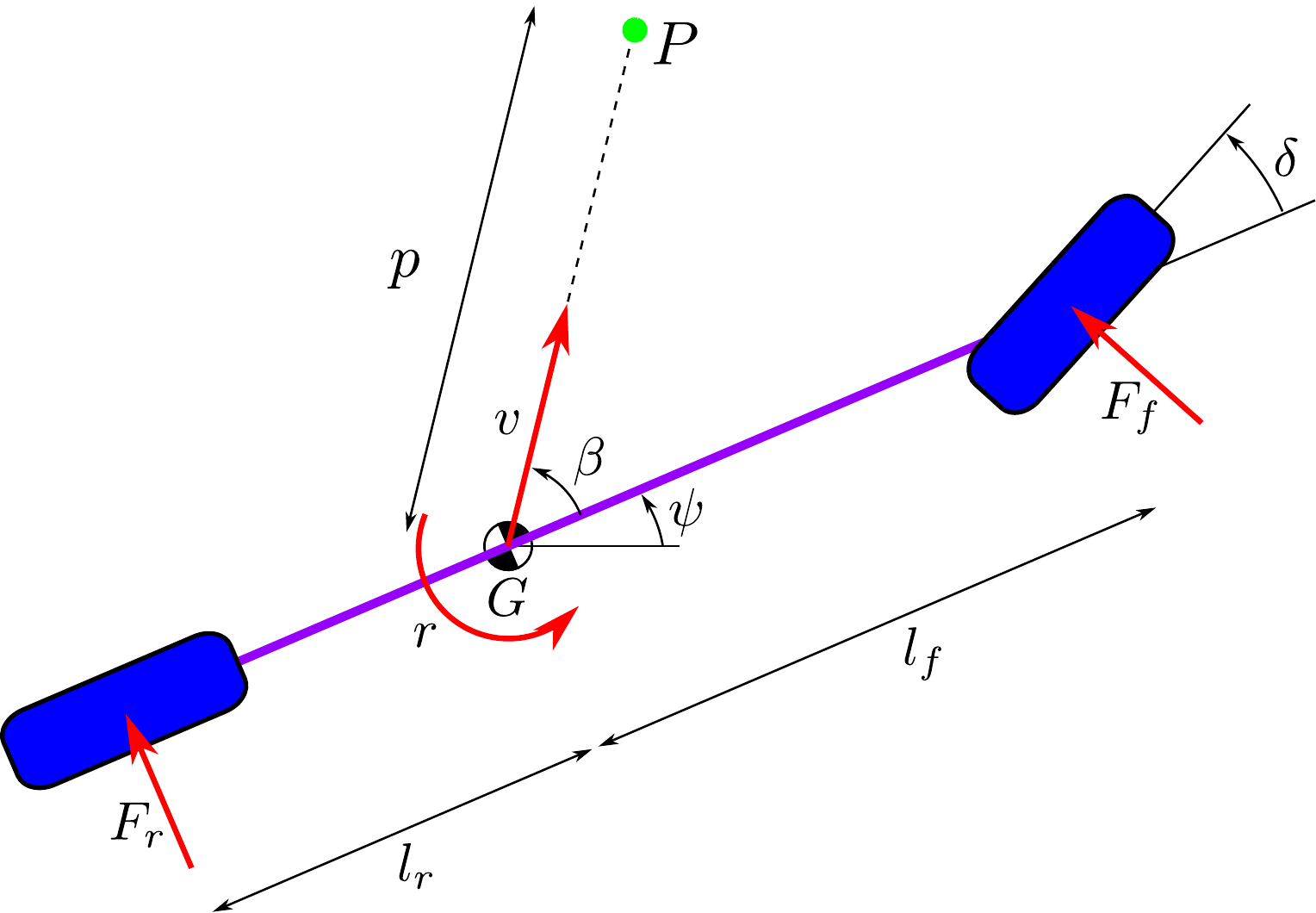}
      \caption{Position of point $\rm{P}$ for the alternative feedback linearisation.	\label{fig:single-track-spaliviero}}
    \end{center}
\end{figure}

The linearising law considered here is inspired by the feedback linearisation developed in~\cite{bib:OrioloEtAl2002} for a unicycle kinematic model, and has been considered and analysed in past works of the authors of this paper, e.g., in~\cite{Spaliviero:Thesis:2016}.\\
In this case point $\rm{P}$, used for devising feedback linearisation, is at a distance $p$ from the center of mass $G$, in the velocity vector direction (Figure~\ref{fig:single-track-spaliviero}). Its coordinates, ${x}_{{\rm\scriptscriptstyle P}}$ and ${y}_{{\rm\scriptscriptstyle P}}$, are given by
\begin{displaymath}
{x}_{{\rm\scriptscriptstyle P}}={x}_{\rm\scriptscriptstyle G}+p\COS{\psi+\beta}\qquad
{y}_{{\rm\scriptscriptstyle P}}={y}_{\rm\scriptscriptstyle G}+p\SIN{\psi+\beta}
\end{displaymath}
and its dynamics is
\begin{equation}
\begin{aligned} \label{eq:P_dynamics_spaliviero}
\dot{x}_{{\rm\scriptscriptstyle P}}&=v\COS{\psi+\beta}-p\left(\dot{\psi}+\dot{\beta}\right)\SIN{\psi+\beta}\\
\dot{y}_{{\rm\scriptscriptstyle P}}&=
v\SIN{\psi+\beta}+p\left(\dot{\psi}+\dot{\beta}\right)\COS{\psi+\beta}
\end{aligned}
\end{equation}
Imposing now suitable reference velocities for point $\rm{P}$, i.e., $v_{{\rm\scriptscriptstyle P},x}$ and $v_{{\rm\scriptscriptstyle P},y}$, by setting
\begin{displaymath}
\dot{x}_{{\rm\scriptscriptstyle P}}=v_{{\rm\scriptscriptstyle P},x}\qquad
\dot{y}_{{\rm\scriptscriptstyle P}}=v_{{\rm\scriptscriptstyle P},y}
\end{displaymath}
and solving equation~\eqref{eq:P_dynamics_spaliviero} with respect to $v$ and $\dot{\psi}+\dot{\beta}$, one obtains
\begin{align*}
v&=v_{{\rm\scriptscriptstyle P},x}\COS{\psi+\beta}+v_{{\rm\scriptscriptstyle P},y}\SIN{\psi+\beta}\\
\dot{\psi}+\dot{\beta}&=\frac{v_{{\rm\scriptscriptstyle P},y}\COS{\psi+\beta}-v_{{\rm\scriptscriptstyle P},x}\SIN{\psi+\beta}}{p}
\end{align*}
Finally, including in the previous equations the dynamics of yaw rate and sideslip, as they are defined by the single-track model~\eqref{eq:singletrack_model}, the following linearising law is obtained
\begin{equation} \label{eq:feedback_lin_spaliviero}
\begin{aligned}
v&=v_{{\rm\scriptscriptstyle P},x}\COS{\psi+\beta}+v_{{\rm\scriptscriptstyle P},y}\SIN{\psi+\beta}\\
\delta&=\frac{m\omega}{C_f} v-\frac{C_r l_r-C_f l_f}{C_f}\frac{r}{v}+\frac{C_r+C_f}{C_f}\beta
\end{aligned}
\end{equation}
where
\begin{displaymath}
\omega=\frac{v_{{\rm\scriptscriptstyle P},y}\COS{\psi+\beta}-v_{{\rm\scriptscriptstyle P},x}\SIN{\psi+\beta}}{p}
\end{displaymath}
A direct comparison of~\eqref{eq:feedback_lin} and~\eqref{eq:feedback_lin_spaliviero} shows that the alternative feedback linearising law, with respect to the one proposed in Section~\ref{sub:sec:feedback_linearization}, is
\begin{itemize}
\item more complex and potentially less robust, as it requires an estimate of vehicle parameters, i.e., mass and cornering stiffness coefficient, that are typically uncertain and potentially time-varying;
\item singular when the vehicle velocity is equal to zero.
\end{itemize}
Some preliminary analysis is conducted in \cite{Spaliviero:Thesis:2016} on the dynamics of the system linearised using~\eqref{eq:feedback_lin_spaliviero}. Such analysis has revealed that, despite the dependence of~\eqref{eq:feedback_lin_spaliviero} upon the cornering stiffness coefficients, the stability properties of the system motions are robust with respect to uncertainties on $C_f$ and $C_r$. However, a further insight allows to highlight some major pitfalls of this approach. More specifically, a bifurcation analysis is conducted using MatCont software tool~\cite{bib:DhoogeEtAl2008}, as done in Section~\ref{sec:structural_stab}.

\begin{figure}[htbp]
	\centering
	\includegraphics[width=0.8\columnwidth]{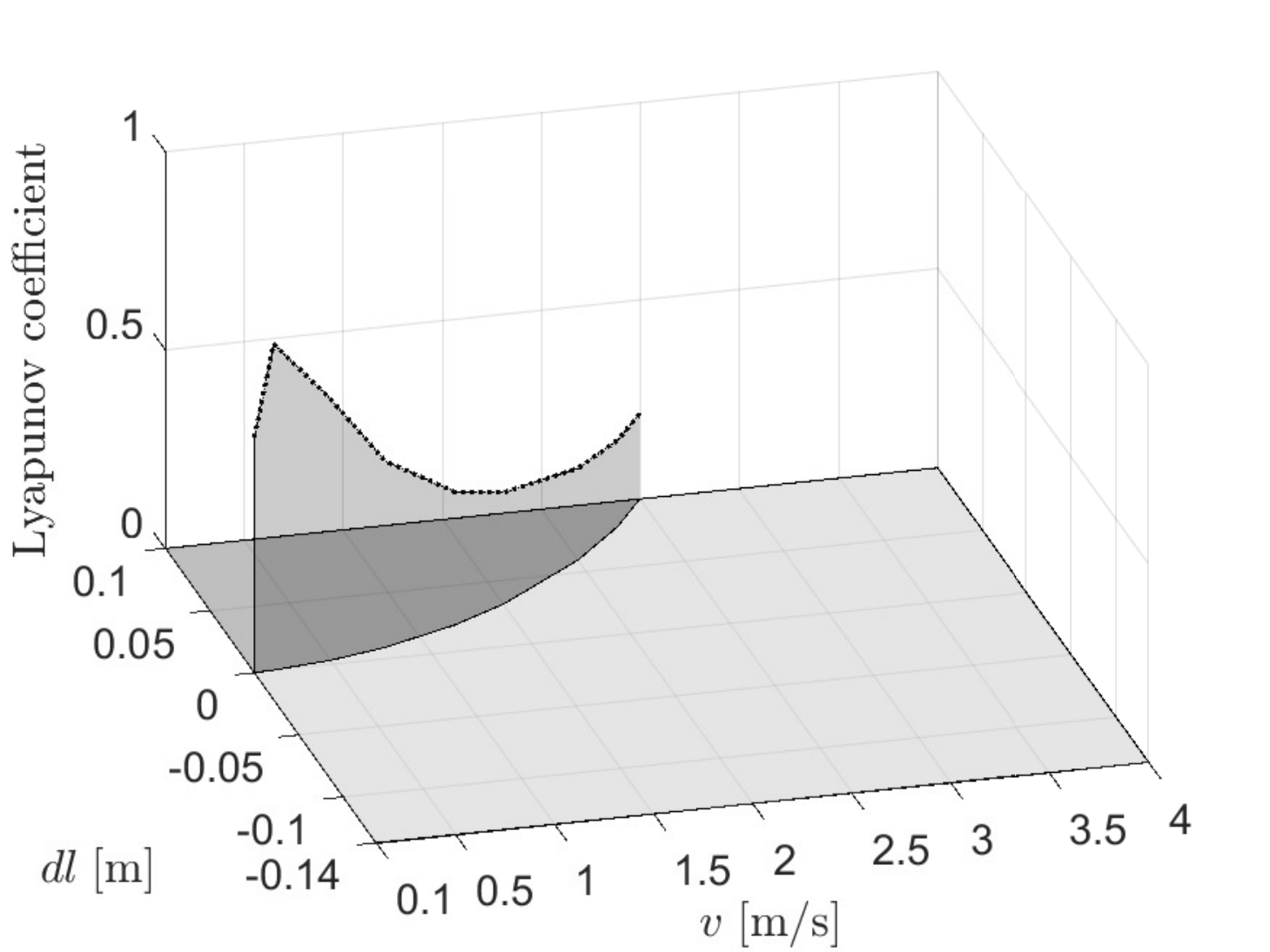} 
	\caption{Stability properties of the selected equilibrium point as a function of $\bar{v}$ and $dl$. Light gray region: asymptotic stability; dark gray region: instability; black solid line: Hopf bifurcation points; dotted black lines: value of the first Lyapunov coefficient for all bifurcation points.}\label{fig:bifurcazSpal}
\end{figure}

\begin{figure}[htbp]
	\centering
	\includegraphics[width=0.8\columnwidth]{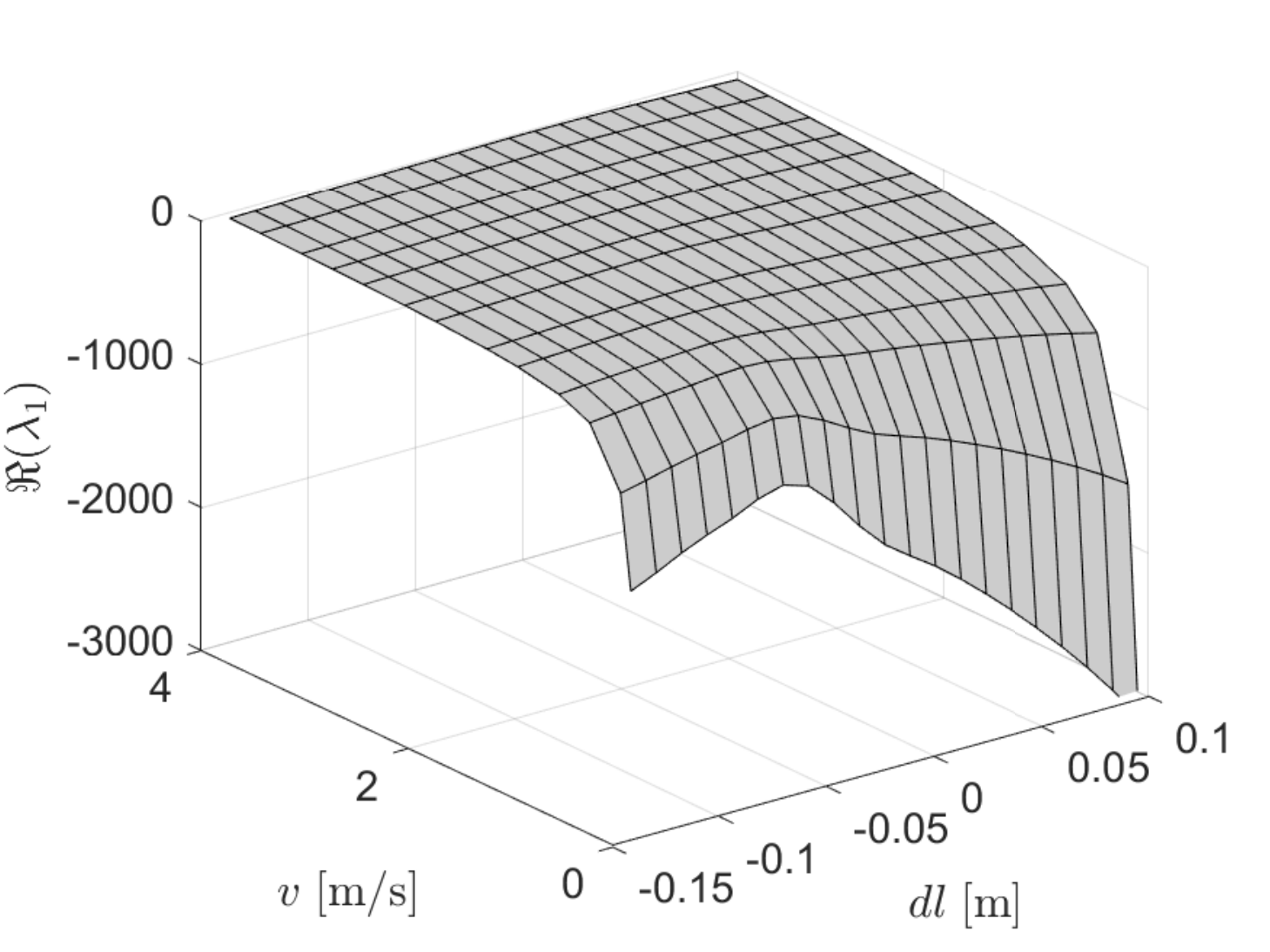}
	\caption{Real part of eigenvalue $\lambda_1$ of the dynamics of $\tilde{\xi}$, linearised around the defined equilibrium value. Light gray is used to define the region in which $\mathcal{R}(\lambda_1)<0$.}\label{fig:l1Spal}
\end{figure}

\begin{figure}[htbp]
	\centering
	\includegraphics[width=0.8\columnwidth]{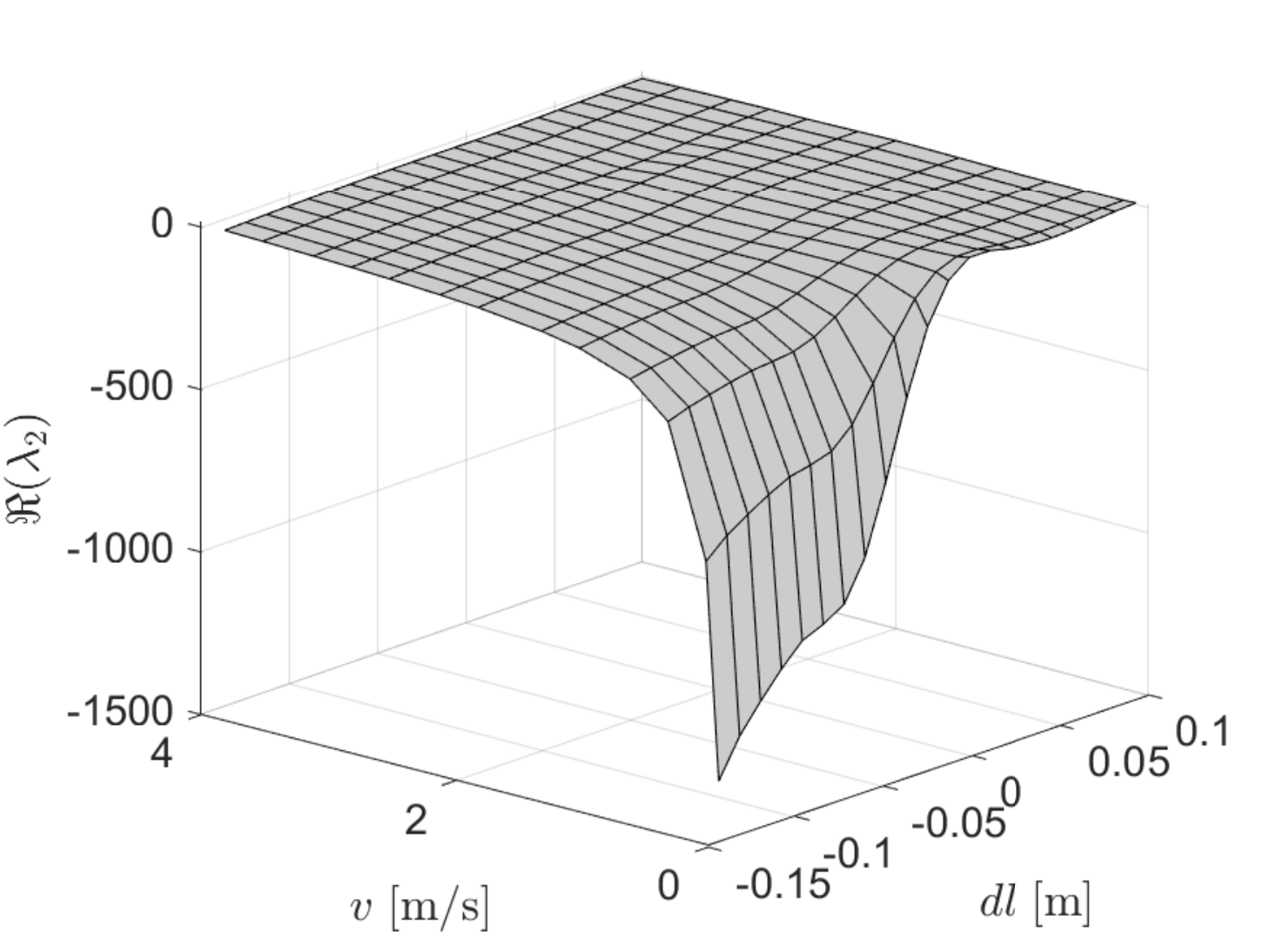}
	\caption{Real part of eigenvalue $\lambda_2$ of the dynamics of $\tilde{\xi}$, linearised around the defined equilibrium value. Light gray is used to define the region in which $\mathcal{R}(\lambda_2)<0$.}\label{fig:l2Spal}
\end{figure}

\begin{figure}[htbp]
	\centering
	\includegraphics[width=0.8\columnwidth]{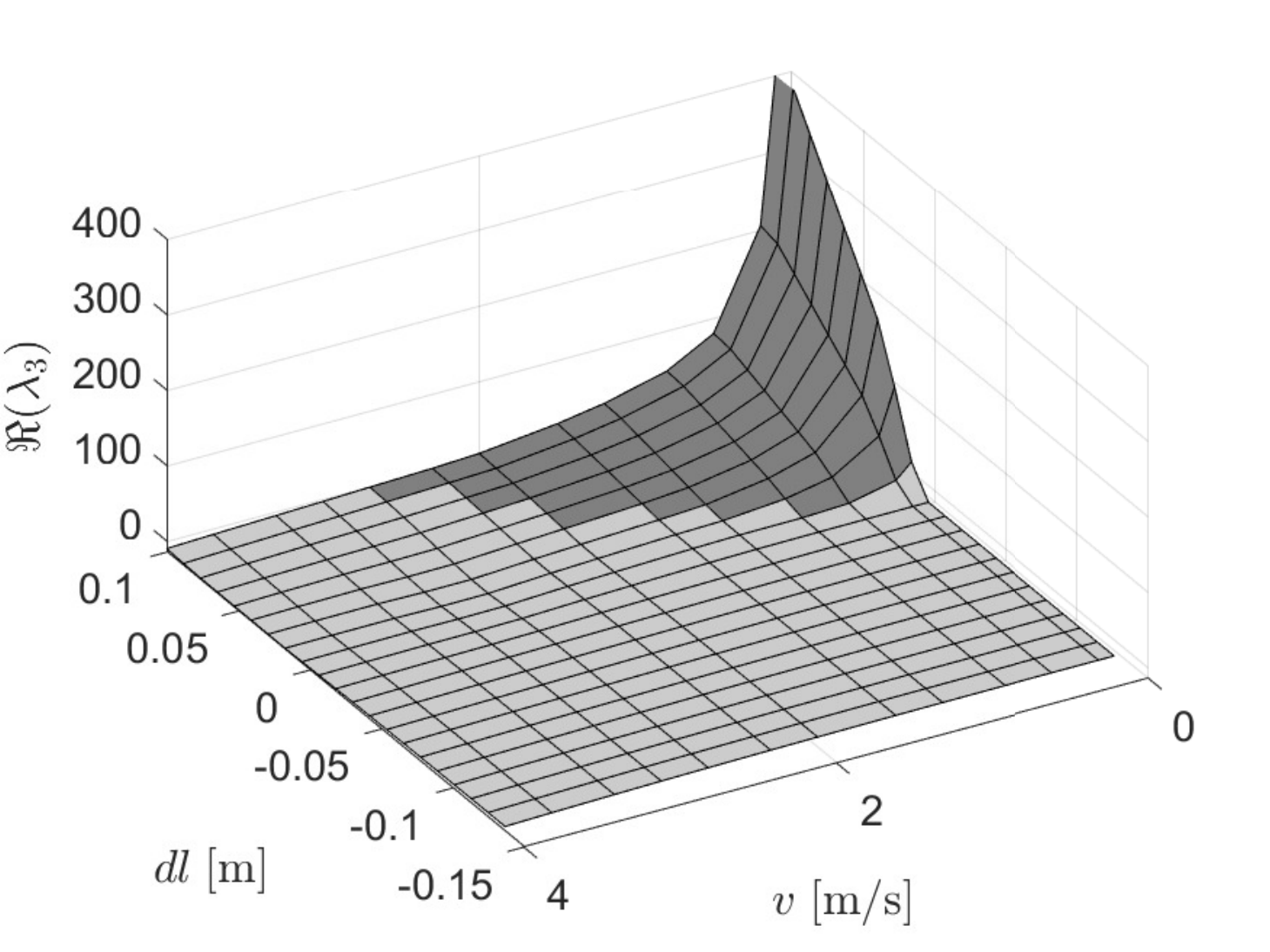}
	\caption{Real part of eigenvalue $\lambda_3$ of the dynamics of $\tilde{\xi}$, linearised around the defined equilibrium value. Light gray is used to define the region in which $\mathcal{R}(\lambda_3)<0$, while dark gray is used to define the region where $\mathcal{R}(\lambda_3)\geq 0$.}\label{fig:l3Spal}
\end{figure}

Define $\tilde{\xi}=\begin{bmatrix}\psi&r&\beta\end{bmatrix}^T$. The dynamics of $\tilde{\xi}$ is governed by~\eqref{eq:singletrack_model},~\eqref{eq:feedback_lin_spaliviero}; in defining the latter, however, we now consider that only the estimate $l_f^{\rm \scriptscriptstyle EST}$ is available, possibly affected by the error $dl=l_f^{\rm \scriptscriptstyle EST}-l_f$. Note that the dynamics of $\tilde{\xi}$ is independent of ${x}_{{\rm\scriptscriptstyle P}}$ and ${y}_{{\rm\scriptscriptstyle P}}$. Therefore, to analyze the properties of the comprehensive system dynamics, the stability analysis described in this section focuses on $\tilde{\xi}$ only.\\
Considering the described setup, $p$ is selected equal to $35\,\mathrm{cm}$, $v_{{\rm\scriptscriptstyle P},x}=\bar{v}\COS{\pi/4}$ and $v_{{\rm\scriptscriptstyle P},y}=\bar{v}\SIN{\pi/4}$. Therefore, the stability of the steady motion $\psi=\pi/4$, $r=0$, and $\beta=0$ is analysed as parameter $dl$ varies in a physically meaningful range, for reasonable values of $\bar{v}$.\\
The results of the bifurcation analysis are summarised in Figure~\ref{fig:bifurcazSpal}, where also the corresponding value of the first Lyapunov coefficient is indicated. 
Subcritical Hopf bifurcations~\cite{bib:DercoleRinaldi2011} are detected at positive values of $dl$. Contrarily to the case of the feedback linearisation algorithm discussed in Section~\ref{sub:sec:feedback_linearization}, these values (especially for small velocities) fall inside the range $[-l_f,l_r]$. In particular, for small velocities, a bifurcation occurs at very small values of $dl$. Indeed, for $v=0.1$, the threshold value for $dl$ is equal to $0.136$ mm, making the approach described here unsuitable in real cases, contrarily to the approach proposed in Section~\ref{sub:sec:feedback_linearization}.\\
For a more thorough insight, in Figures~\ref{fig:l1Spal}-\ref{fig:l3Spal} the values of the real parts of the four eigenvalues of the dynamics of $\tilde{\xi}$, linearised around the defined equilibrium value, are shown. This clearly shows the dependence of the eigenvalues, and especially of the most critical one (i.e., $\lambda_3$), upon the uncertainty, for all considered velocity conditions. 
 

\bibliographystyle{IEEEtran}
\bibliography{IEEEabrv,FBLIN_CSS-L}
	
\end{document}